\documentclass[%
reprint,
superscriptaddress,
 amsmath,amssymb,
prb,
]{revtex4-2}

\usepackage[main=english]{babel}
\usepackage{graphicx}
\usepackage{dcolumn}
\usepackage{bm}
\usepackage{todonotes}
\usepackage{hyperref}
\usepackage{subcaption}
\usepackage{verbatim}
\usepackage[normalem]{ulem}
\usepackage{dsfont}
\usepackage{cancel}
\usepackage{amsmath}
\usepackage{physics}
\newcommand{\adagop}[1]{\hat{a}_{#1}^\dagger}
\newcommand{\aop}[1]{\hat{a}_{#1}}
\newcommand{\alphadagop}[1]{\hat{\alpha}_{#1}^\dagger}
\newcommand{\alphaop}[1]{\hat{\alpha}_{#1}}

\newcommand{\bdagop}[1]{\hat{b}_{#1}^\dagger}
\newcommand{\bop}[1]{\hat{b}_{#1}}
\newcommand{\betadagop}[1]{\hat{\beta}_{#1}^\dagger}
\newcommand{\betaop}[1]{\hat{\beta}_{#1}}
\newcommand{\rhodp}[0]{\rho_{d\phi}}

\newcommand{\rhotimederivative}[0]{\dot{\rho}(t)}
\newcommand{\rhot}[0]{\rho(t)}
\begin{document}

\preprint{APS/123-QED}

\title{Universal Quantum Gate Set for  Multi-Mode Schrödinger Cat States Stabilized by Non-Local Dissipation Engineering}

\author{Jesper Lind-Olsen}
 \affiliation{Department of Physics, University of Oslo, Oslo, Norway}
\affiliation{Norwegian National Security Authority, Kolsås, Norway}

\author{Jonas Lidal}
\affiliation{Norwegian National Security Authority, Kolsås, Norway}

\author{Tron Omland}
\affiliation{Norwegian National Security Authority, Kolsås, Norway}
\affiliation{Department of Mathematics, University of Oslo, Oslo, Norway}

\author{Joakim Bergli}
\affiliation{Department of Physics, University of Oslo, Oslo, Norway}

\date{\today}
\begin{abstract}
Schrödinger cat states provide a hardware-efficient platform for bosonic quantum error correction by encoding logical information in protected manifolds of harmonic oscillators. While previous work has demonstrated the dissipative stabilization of multi-mode Schrödinger cat states as robust quantum memories, a framework for universal quantum computation has remained unavailable. Here we work towards this  by introducing a universal gate set for dissipatively stabilized multi-mode cat qubits as a necessary resource for computation. Using a chain of Kerr non-linear oscillators coupled through engineered non-local dissipation and an effective low-dimensional description, we show how  single-qubit control can be achieved through arbitrary rotation around the $X$-axis and $\pi/2$-rotation around the $Z$-axis. We further show how coupling two such stabilized arrays through just one oscillator on each respective array enables coherent entangling operations through implementation of the $XX_{\pi/2}$ gate. Numerical simulations demonstrate high-fidelity gate dynamics and entanglement generation under realistic parameters. Finally, we analyze the effects of induced and intrinsic photon loss, disorder, and the validity regime of the effective low-dimensional theory. Our results establish dissipatively stabilized multi-mode Schrödinger cat states as a potential architecture for universal bosonic quantum computation.

\end{abstract}

\maketitle

\section{\label{sec:Introduction}Introduction}
Quantum computing has long promised a host of intriguing properties~\cite{cheng_noisy_2023, huang_vast_2025}. These include exponential speed up of computations~\cite{pokharel_demonstration_2023, shor_algorithms_1994, shor_polynomial-time_1997, singkanipa_demonstration_2025}, intrinsically secure communication~\cite{hacker2019c, jeong2001b, vanenk2001c} and a new paradigm for simulation~\cite{altman_quantum_2021}. It has become increasingly clear through both theoretical and experimental breakthroughs that the main challenge is still the inherent fragility of quantum states~\cite{bharti_noisy_2022, cheng_noisy_2023}. The effects of noise, causing decoherence and decreased fidelity still plague the field. Barring the discovery of completely fault-tolerant components, the need for detection and reliable correction of errors is of paramount importance~\cite{terhal_quantum_2015}.

One route towards this goal is that of using Schrödinger cat states, superpositions of macroscopically distinct quantum states~\cite{wang2016c, liao_generation_2016, he_generation_2023}. Due to their rotational symmetry , these states are considered a viable option for quantum computing. Rotationally symmetric bosonic error correction codes using such states have been proposed~\cite{grimsmo_quantum_2020}, while the states themselves have been prepared experimentally with high fidelity in several setups, including microwave cavities and superconducting circuits~\cite{grimm2020d, ofek2016c}. Bosonic qubits, such as Schrödinger cat based models, are seen as a hardware-efficient method of encoding quantum information as the high-dimensional Hilbert space of a single harmonic oscillator provides the redundancy to perform quantum error correction~\cite{gottesman_encoding_2001}.

Earlier research has focused primarily on single-mode Schrödinger cat states~\cite{sanders2012a, gilles1994b,grimm2020d}, however, recent progress has seen demonstrations of their multi-mode counterpart~\cite{zapletal_stabilization_2022}. By entangling multiple bosonic modes into structured superpositions, one can bias the resulting states to be more tolerant to specific errors, while simultaneously being able to encode the logical information non-locally over a spatially distributed setup. Beyond only computation, these states also exhibit enhanced properties in the field of quantum sensing and metrology~\cite{munro2002c,ralph2002c,zhang2013a, joo2011c}, making them an interesting topic of study across several related fields.

In this article we will extend previous work done to dissipatively engineer multi-mode Schrödinger cat states. Previous results have shown stabilized quantum memory and the single qubit bit-flip gate. We will build on this to include the necessary resources for a universal gate set by which to perform  logical operations on  information encoded in these states~\cite{goto_universal_2016, mirrahimi_dynamically_2014}. We show that we can implement a second one-qubit rotation, which allows for arbitrary one-qubit operations.  We also show that we can extend to multiple qubits and perform entangling operations between two multi-mode Schrödinger cat qubits with direct interaction of only two modes.

\section{\label{sec:Model}Model}
\begin{figure}
\includegraphics[width=0.9\linewidth]{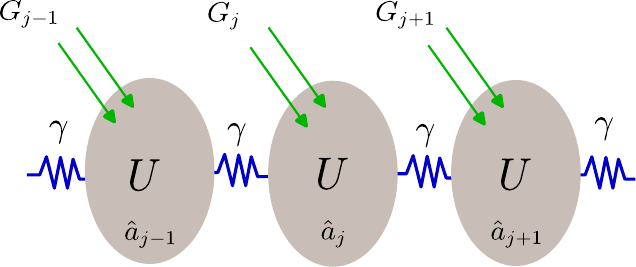} 
\caption{An illustration of the setup. A set of Kerr anharmonic oscillators with Kerr non-linearity $U$ and driven by a two-photon pump with amplitude $G$ coupled through a lossy intermediate reservoir with strength $\gamma$. The whole system has periodic boundary conditions.}
\label{fig:model}
\end{figure}
\
We consider the model introduced by Zapletal \textit{et al.}~\cite{zapletal_stabilization_2022}; namely a 1D chain of anharmonic oscillators (AHO) with Kerr-type nonlinearities, two-photon pumping at twice the oscillator frequency and periodic boundary conditions. The Hamiltonian of this system in the lab frame is
\begin{equation}
    \hat{H} = \omega_c\sum_j \adagop{j}\aop{j} + U\sum_j(\adagop{j})^2(\aop{j})^2 + \sum_j\big[G_j(\adagop{j})^2e^{-i\omega_dt}+h.c.\big], \label{Hamiltonian real space lab frame}
\end{equation}
where $\adagop{j}$ ($\aop{j}$) are bosonic raising (lowering) operators. $U$ is the amplitude of the Kerr non-linearity, while $G_j = \abs{G}e^{i\theta_j}$ is the two-photon pumping amplitude.  $\omega_{c}$ is the cavity frequency, while $\omega_d$ is the driving frequency. We couple these AHOs to each other dissipatively. The GKS-Lindblad equation (with units where $\hbar =1$) is

\begin{equation}
    \frac{\partial \rho(t)}{\partial t} = -i\comm{\hat{H}}{\rho(t)} + \mathcal{L}_d\rho(t) \label{lindblad equation}.
\end{equation}
The particular form of dissipative coupling used in this case is described by the superoperator
\begin{equation}
    \mathcal{L}_d \rho = \sum_{j}\gamma D[\aop{j}-e^{i\phi}\aop{j+1}]\rho, \label{dissipator}
\end{equation}
which we will refer to informally as the dissipator. Here, $D[\hat{O}]\rho = \hat{O}\rho\hat{O}^\dagger -\frac{1}{2}\acomm{\hat{O}^\dagger\hat{O}}{\rho}$. We also enforce periodic boundary conditions, \textit{i.e.} $\aop{j}=\aop{j+N}$ for a chain consisting of $N$ oscillators. This form of dissipative coupling can be engineered with  intermediate lossy cavities with decay rate $\kappa_l$ coupled to  two neighboring target cavities with  hopping amplitudes $J > 0$ and $-Je^{i\phi}$, respectively~\cite{li_dissipation-induced_2021, metelmann_nonreciprocal_2015, mamaev_dissipative_2018}. . For $\kappa_l\gg J$, formally tracing out the reservoir cavities leads to effective coupling on the target cavities of Eq. \eqref{dissipator}, with $\gamma=4J^2/\kappa_l$ in the resonant frame. In this setting, $\phi$ is the relative phase difference between the coupling of two neighboring target cavities to the  intermediate reservoir cavity. The setup is shown schematically in Fig. \ref{fig:model}. Taking the driving frequency $\omega_d$ to be twice the oscillator frequency, \textit{i.e.} $\omega_d=2\omega_c$ and transforming the Hamiltonian to the frame rotating at oscillator frequency by $\hat{U} = \exp{(-i\omega_c t\sum_j\adagop{j}\aop{j})}$ we can write the Hamiltonian as
\begin{equation}
    \hat{H} = U\sum_j(\adagop{j})^2(\aop{j})^2 + \sum_j\big[G_j(\adagop{j})^2+h.c.\big]. \label{Hamiltonian real space}
\end{equation}
We take the pump phases $\theta_j=-2j\phi$ as in~\cite{zapletal_stabilization_2022}.
Due to the translational symmetry in this problem, we make use of the plane wave basis 
\begin{equation}
    \alphaop{k}=\frac{1}{\sqrt{N}}\sum_j e^{ikj}\aop{j} \label{fourier transform op}
\end{equation}
with $k=n\frac{2\pi}{N}$ for $n\in\{0,\dots,N-1\}$, and $\phi \in (2\pi/N)\mathds{Z}$ so $k=\phi$ is an allowed mode, and transfer the problem to Fourier space.   Inserting the inverse of Eq. \eqref{fourier transform op} into Eq. \eqref{dissipator} and resolving the sums, the dissipator now takes the non-interacting form
\begin{equation}
    \mathcal{L}_d \rho = \sum_k \gamma_k D[\alphaop{k}]\rho, \label{dissipator k space}
\end{equation}

where
\begin{equation}
    \gamma_k = 2\gamma\big[1-\cos{(k-\phi)}\big] \label{gamma k}.
\end{equation}
The interpretation of this term is that we have a non-uniform one-photon loss over the non-local $k$-modes shown in Eq. \eqref{dissipator k space}. From Eq. \eqref{gamma k}, we see that for the particular mode $k=\phi$, the loss rate vanishes, leaving this mode unaffected by dissipation. We therefore refer to this as a dark mode. Furthermore, we will also consider the so called dark states of the system - eigenstates of the Hamiltonian that lie on the dark mode and are therefore unaffected by dissipation. An alternative model with the same results was also presented in the same paper by Zapletal \textit{et al.}~\cite{zapletal_stabilization_2022}, where the Kerr nonlinearity is replaced by local
two-photon loss.

\subsection{Decoherence Free Subspace and Dark States\label{DFS}}
As shown in~\cite{zapletal_stabilization_2022},  Eq. \eqref{lindblad equation} has an analytical steady state  in reciprocal space, for states satisfying the two criteria
\begin{align}
    \hat{H}\ket{\Psi} &= \epsilon\ket{\Psi} \label{schrodinger eq}\\
    \alphaop{k} \ket{\Psi} &= 0\text{ } \forall \text{ }k\neq \phi \label{criteria operators}
\end{align}
which follow from Eqs. \eqref{dissipator k space} and \eqref{lindblad equation}.  Equations \eqref{schrodinger eq} and \eqref{criteria operators} are sufficient to enforce that the RHS of Eq. \eqref{lindblad equation} equals zero. This gives $\dot{\rho}(t) = 0$, and we solve for the steady state. Equation \eqref{criteria operators} enforces a solution of the following form
\begin{equation}
    \ket{\Psi} = \bigg(\bigotimes_{k\neq \phi} \ket{0}_{}\bigg) \bigotimes\ket{\psi}_\phi,\label{steady state form}
\end{equation}
while solving equation Eq. \eqref{schrodinger eq} with the eigenvalue $\epsilon = G\zeta^2$, where $\zeta =  i\sqrt{\frac{NG}{U}}$, gives the lowest energy analytical solution to the steady state as a state lying in the span of 
\begin{equation}
    \ket{\Psi_{}}  \propto \bigg( \bigotimes_{k\neq \phi} \ket{0} \bigg)\bigotimes\bigg(\ket{\zeta} \pm \ket{-\zeta} \bigg).
\end{equation}
The effect of the engineered dissipation is to pick out a single-mode which can be populated in the infinite time limit, and any superpositions of the two coherent states $\ket{\pm\zeta}$ are dark states. We choose to use the basis of Schrödinger cat states
\begin{equation}
    \ket{C_\pm} = \mathcal{N_\pm}\big(\ket{\zeta} \pm \ket{-\zeta}\big)  \label{cat states def}
\end{equation}
to describe the dark states. $\mathcal{N}_\pm = [2(1\pm e^{-2\abs{\zeta}^2})]^{-1/2}$ is a normalization factor.
Transforming back to the local basis, the solution now takes the form
\begin{equation}
    \ket{\Psi_{}} = \mathcal{N_\pm}\bigg(\bigotimes_{j=1}^N\ket{\zeta_j}_j \pm \bigotimes_{j=1}^N\ket{-\zeta_j}_j\bigg),
\end{equation}
with $\zeta_j = \frac{1}{\sqrt{N}}\zeta e^{-ij\phi}$. This is a truly multi-mode entangled Schrödinger cat state, and, in contrast to the product state $\ket{\Psi} =\bigotimes_{j=1}^N \big(\ket{\zeta_j} \pm \ket{-\zeta_j}\big), $
the truly multi-mode Schrödinger cat state offers the possibility of an increased error threshold through the effects of error biasing~\cite{lescanne_exponential_2020,puri_engineering_2017, mirrahimi_dynamically_2014}, while also having multi-mode entanglement and being a possible resource state for quantum sensing and metrology~\cite{joo2011c, munro2002c}.

\subsubsection{The Logical Subspace} \label{Logical subspace}
To build our logical subspace we will make use of the Zeno limit to approximate the system into an effective model. Named after the Zeno paradox, the Zeno limit is an approximation whereby projective measurements on parts of the system restrict any time evolution of the quantum state in that subsystem~\cite{popkov_effective_2018}. This can be utilized by shrinking the Hilbert space for which there is time evolution. In the case at hand, there are no projective measurements; however, in the strong dissipation regime, this dissipation will act as a projective measurement. Any excitations on modes $k\neq \phi$ will subsequently de-excite into the vacuum, rendering these modes stationary when the limit of strong dissipation is applied. Thus, as we will formally introduce in Sec.\ref{zeno limit coupled chains}, we may project the instantaneous state onto the vacuum in all the ``stationary" modes and trace out these, leaving only one evolving mode. As we will also see when we introduce a coupling to a second, identical system, the same formalism can describe a two-mode model. 

To describe the logical operations we need to define a logical subspace on which these operators will act, which we will also refer to as the decoherence free subspace (DFS). We consider that the logical subspace is the following 
\begin{equation}
    V_\phi  = \text{span}\big(\ket{C_+}_\phi, \ket{C_-}_\phi\big),\label{logical subspace}
\end{equation} 
namely states lying in the span of two basis states, the odd and even Schrödinger cat states, on the dark mode $\phi$. We therefore consider our logical basis states $\ket{0/1}_L = \ket{C_\pm}_\phi$ in addition to the complementary basis $\ket{\pm}_L = 1/\sqrt{2}(\ket{C_+}_\phi \pm \ket{C_-}_\phi)$.  To build a universal gate set, we need to be able to perform arbitrary one-qubit rotations, together with one entangling gate between two qubits~\cite{bremner_practical_2002}.

Defining a logical subspace as stated in Eq. \eqref{logical subspace}, a one-photon drive, $H_\eta = \eta \left( e^{ij' \phi} \hat{a}_{j'} + e^{-i j' \phi} \hat{a}_{j'}^\dagger \right)$, applied to a single oscillator, $j'$, can be used to implement arbitrary rotations around the $X$-axis~\cite{zapletal_stabilization_2022}. For single-mode Schrödinger cat qubits, a $\pi/2$ rotation around the $Z$-axis can be implemented by turning off two-photon effects and including the Kerr non-linearity $-\chi_{K} (\adagop{}\aop{})^2$~\cite{mirrahimi_dynamically_2014}. In Appendix \ref{appendix rotations} we show that by leveraging the same mechanics, namely turning off two photon effects and replacing them by a self-Kerr effect on each of the constituents on the chain of dissipatively coupled oscillators we can achieve effective one-mode dynamics replicating this term in the Zeno limit, performing a $\pi/2$ rotation around the $Z$-axis in the protected subspace. These two operations are sufficient for arbitrary one-qubit rotations~\cite{mirrahimi_dynamically_2014}, which we show in Appendix \ref{appendix rotations}. In the next section we go through the details of how one couples two of these qubits to each other and perform an entangling gate between them, which concludes a universal gate set.

\subsection{The Two Qubit Model}
\begin{figure}[h]

\includegraphics[width=0.9\linewidth, height=6cm]{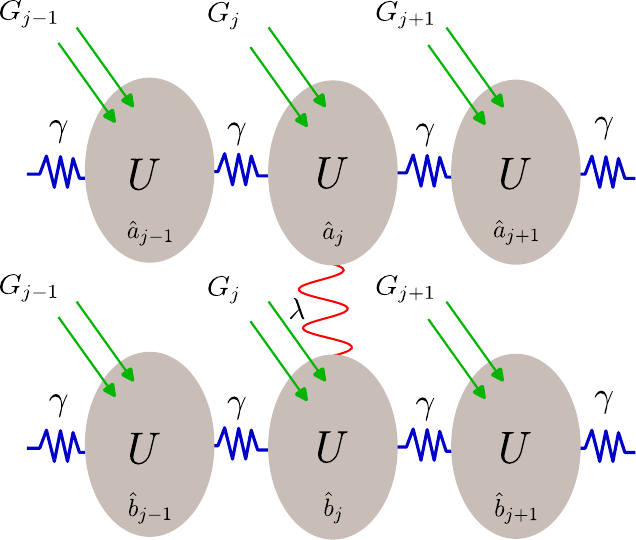}
\caption{The two qubit model includes the coupling Hamiltonian $\hat{H}_{BS} = \lambda(\aop{j}\bdagop{j} + \adagop{j}\bop{j})$.}
\label{fig:two qubit illustration}

\end{figure}
We couple two arrays to each other at a single resonator on each chain using the coupling Hamiltonian
\begin{equation}
    \hat{H}_{BS} = \lambda(\aop{l}\bdagop{l'} + \adagop{l}\bop{l'}). \label{H BS}
\end{equation}
Here $\aop{j}^{(\dagger)} (\bop{j}^{(\dagger)})$ denote bosonic operators on the A(B) subsystems, respectively, and $l, l'$ are two particular AHOs from each chain. We show that using only one such coupling in the Zeno limit gives the effective interaction needed in the $\phi$ modes of each constituent system to induce the same Hamiltonian on the protected modes of each of the arrays. A schematic figure of the setup is shown in Fig. \ref{fig:two qubit illustration}, where $l = l' = j$ and the strength of the coupling is denoted by $\lambda$, and is assumed to be tunable.

\subsubsection{Deriving an Effective Model} \label{deriving eff model}

Using the aforementioned model, we can derive an analytical approximation for the effective two mode dynamics of the coupled chains, using the Zeno limit. Each mode supports one Schrödinger cat qubit, and the coupling will allow for entanglement and two qubit gates. Our starting point is the Hamiltonian
\begin{equation}
    \hat{H} = \hat{H}_A + \hat{H}_B + \hat{H}_{BS}
\end{equation}
where  $\hat{H}_{A}$ and $\hat{H}_{B}$ are the Hamiltonians from Eq. \eqref{Hamiltonian real space} for each of the two chains labeled by A and B, with the respective bosonic operators  $\aop{j}^{(\dagger)}$ and $\bop{j}^{(\dagger)}$ corresponding to the labeling. $H_{BS}$ is the beamsplitter Hamiltonian from Eq. \eqref{H BS}. In addition, both chains have non-local dissipation as given in Eq. \eqref{dissipator}, where $\aop{j}/\adagop{j}$ is replaced by $\bop{j}/\bdagop{j}$ for the second chain. Both the phase difference $\phi$ and the dissipation rate $\gamma$ are assumed to be identical on both setups. We will also for the remainder set $l=l'=1$ for some arbitrary labeling of the ``first" AHO on each chain. As opposed to the case of cat-state preparation proposed by Zapletal \textit{et al.} in \cite{zapletal_stabilization_2022}, where parity was an approximately conserved quantity - exact in the Zeno limit - the beamsplitter explicitly breaks this, and preparation of cats must be done on decoupled chains ($\lambda=0$). Inspired by the analysis on a single chain, we transform the problem to Fourier space. The Fourier transformed Hamiltonian in the stationary frame reads
\begin{widetext}

\begin{align}
    \begin{split}
    \hat{H}_k &= \frac{U}{N}\sum_{k_1,\dots ,k_4} \alphadagop{k_1}\alphadagop{k_2}\alphaop{k_3}\alphaop{k_4}\delta_{k_1+k_2,k_3+k_4}+ \frac{U}{N}\sum_{k_1,\dots, k_4} \betadagop{k_1}\betadagop{k_2}\betaop{k_3}\betaop{k_4}\delta_{k_1+k_2,k_3+k_4}   \\
    &+ G\sum_{k_1}\bigg[ \alphadagop{k_1}\alphadagop{2\phi-k_1} + h.c. \bigg] + G\sum_{k_1}\bigg[ \betadagop{k_1}\betadagop{2\phi-k_1} + h.c. \bigg]  +\frac{\lambda}{N}\sum_{k_1,k_2} \bigg[ e^{i(k_1-k_2)}\alphadagop{k_1}\betaop{k_2} + h.c.\bigg].
    \end{split} \label{k space hamiltonian}
\end{align}
\end{widetext}
Here, $N=N_A = N_B$ is the number of AHOs in each chain, which we take to be equal. $\betaop{k}$ is defined in the same way as $\alphaop{k}$, by exchanging $\aop{j}$ for $\bop{j}$ in Eq. \eqref{fourier transform op}. Including the Fourier transformed dissipative term Eq. \eqref{dissipator k space} for both chains, with $\alphaop{k}$ swapped for $\betaop{k}$ for the B subsystem, the Lindblad equation is 
\begin{align}
    \begin{split}
    \rhotimederivative = -i\comm{\hat{H}_k}{\rhot} + \sum_k\gamma_k \bigg(D\big[\alphaop{k}\big] + D\big[\betaop{k}\big]\bigg)\rhot. \label{two mode lindblad}
    \end{split}
\end{align}

\subsubsection{Zeno Limit of Coupled Chains} \label{zeno limit coupled chains}

Inspired by Zapletal \textit{et al.}'s analysis on a single chain~\cite{zapletal_stabilization_2022}, we seek to derive the restricted dynamics in the quantum Zeno limit. We will use an ansatz for the state where the dark modes are free to evolve under effective dynamics, while the remaining modes are locked in the vacuum. We write this state as

\begin{equation}
    \rho(t) = \bigotimes_{k_1 \neq \phi} \ket{0}_{k_1}\bra{0}\bigotimes_{k_2\neq \phi}\ket{0}_{k_2}\bra{0}\otimes \rho_{\phi\phi}(t) \equiv \rho_{d}\otimes \rho_{\phi\phi}(t) \label{state ansatz},
\end{equation}
where the dynamics of $\rho_{\phi\phi}(t)$ are governed by an effective two-mode evolution arising in the Zeno limit on the two non-dissipative modes - one on each chain. We now derive the effective two-mode system. We rescale the time variable in the Lindblad equation $\tau = \gamma t$. Under this transformation we may rewrite Eq. \eqref{two mode lindblad} to get
\begin{equation}
    \frac{\partial \rho(\tau)}{\partial\tau} = (\mathcal{L}_d + \mathcal{K})\rho(\tau) \label{rescaled meq},
\end{equation}
where we have defined 
\begin{align}
    \mathcal{L}_d \rho(\tau) &= \sum_{k\neq \phi} \frac{\gamma_k}{\gamma} \bigg(D\big[\alphaop{k}\big] + D\big[\betaop{k}\big]\bigg) \rho(\tau) \\
    \mathcal{K}[\rho(\tau)] &= -\frac{i}{\gamma}\comm{\hat{H}_k}{\rho(\tau)}.
\end{align}
We leave it implicit that $\alphaop{k} / \betaop{k}$ act on distinct parts of the Hilbert space, while using the same label. A general solution to Eq. \eqref{rescaled meq} exists in terms of the propagator $U(\tau)$ as
\begin{equation}
    \rho(\tau) = U(\tau)\rho(0),
\end{equation}
where $U(\tau)$ is defined self-consistently through~\cite{breuer_theory_2007} 
\begin{equation}
    U(\tau) = e^{\mathcal{L}_d\tau} \bigg(1+\int_0^\tau \mathrm{d}\tau_1 e^{-\mathcal{L}_d\tau_1}\mathcal{K}U(\tau_1) \bigg). 
\end{equation}
The self-consistent equation for $U(\tau)$ can be solved iteratively, giving rise to a Dyson-series expansion 
\begin{align}
    \begin{split} 
    U(\tau) &= e^{\mathcal{L}_d\tau} \bigg(1+\int_0^\tau \mathrm{d}\tau_1 e^{-\mathcal{L}_d\tau_1}\mathcal{K} e^{\mathcal{L}_d\tau_1} \\
    &+ \int_0^\tau \mathrm{d}\tau_1 e^{-\mathcal{L}_d\tau_1}\mathcal{K} e^{\mathcal{L}_d\tau_1}\int_0^{\tau_1} \mathrm{d}\tau_2 e^{-\mathcal{L}_d\tau_2}\mathcal{K} e^{\mathcal{L}_d\tau_2} +\dots \bigg), \label{Dyson series}
    \end{split}
\end{align}
where we, in line with the reasoning for the Zeno limit, assume $\gamma \gg U, G, \lambda$, allowing us to truncate the expansion at the second order in $\mathcal{K} \propto 1/\gamma$. Going to a higher order in the Dyson series would entail extra incursions into the bright modes. This coupling would be of order $\sim U/N$, and would be suppressed by the bright mode gap $\gamma_g = 2\gamma[1-\cos{(k_{bright}-\phi)}]$. The neglected terms are therefore at leading order $\sim U/\gamma_gN$ \textit{relative} to the retained terms. Inserting the parameters used for the numerical results of Fig.  \ref{fig:Dynamics:fidelity}, this is of the order $\sim 10^{-3}$, with an increase in $\gamma$ further suppressing this. Additional comments on truncation and residual error are left to Appendix \ref{Full dynamics} where the full Lindbladian dynamics are compared to the Zeno-reduced dynamics.

Denote by $\Tr_d$ and $\Tr_\phi$ the trace over all modes $k\neq \phi$ and over only $k=\phi$, respectively. Here we have suppressed the double indices and leave it implicit that the trace is over modes on both chains.
Since the two modes with $k =\phi$ do not undergo any dissipation, $(\Tr_d\mathcal{L}_d)\rho_{\phi\phi} = \rho_{\phi\phi}$, where we have defined
\begin{equation}
    \Tr_d\rho \equiv \rho_{\phi\phi} 
\end{equation} 
to be the reduced density matrix for the two $\phi$ modes. The unique steady state in the subspace $d$, $(\Tr_\phi\mathcal{L}_d)\rho_d$, is the vacuum state $\rho_d = \bigotimes_{k\neq\phi} \ket{0}_k\bra{0}$.  The projection onto the kernel of $\mathcal{L}_d$ is the superoperator~\cite{popkov_effective_2018}
\begin{equation}
    P_d \equiv  \lim_{\tau\to\infty}\exp{(\mathcal{L}_d\tau)}
\end{equation}
which acts on operators of the full system. It satisfies $P_d^2=P_d$ and
\begin{equation}
    e^{\tau \mathcal{L}_d} P_d = P_de^{\tau\mathcal{L}_d} = P_d,\label{properties projector}
\end{equation}
so it projects onto the kernel of $\mathcal{L}_d$ along its range. The action of the projection operator on an arbitrary operator is 
\begin{equation}
    P_d \hat{X} = \rho_d \otimes \Tr_d\hat{X}.
\end{equation}
To obtain an equation for the reduced propagator $U(\tau)_{\phi\phi} = P_d U(\tau)P_d$ we project Eq. \eqref{Dyson series} onto the reduced subspace. Using the properties previously described for the projectors  and truncating at second order in $\mathcal{K}$, the reduced propagator satisfies
\begin{align}
    \begin{split}
        U(\tau)_{\phi\phi} &= P_d + \tau P_d \mathcal{K}P_d\\
        &+ P_d\mathcal{K} \int_0^\tau \mathrm{d}\tau_1 \int_0^{\tau_1}\mathrm{d}\tau_2  e^{\mathcal{L}_d(\tau_1-\tau_2)}\mathcal{K} P_d. \label{red propagator}
    \end{split}
\end{align}
Solving the integrals, formally doing the projections, transforming back to unscaled time and taking the equation back to differential form we get (Appendix \ref{Appendix Zeno})

\begin{align}
    \begin{split}
        \frac{\partial \rho_{\phi\phi}(t)}{\partial t}  &= -i\comm{\hat{H}_\phi}{\rho_{\phi\phi}(t)}
        + \chi \sum_{\hat{z}_\phi\in\{\alphaop{\phi},\betaop{\phi}\}} D[\hat{z}_\phi]\rho_{\phi\phi}(t)
        \\ &+ \Gamma\sum_{\hat{z}_\phi\in\{\alphaop{\phi},\betaop{\phi}\}} D[\hat{z}^2_\phi-\zeta^2]\rho_{\phi\phi}(t)\label{Final lindblad eq}.
    \end{split}
\end{align}
Here, the effective Hamiltonian is 
\begin{align}
    \hat{H}_\phi &= \frac{U}{N}(\alphadagop{\phi})^2(\alphaop{\phi})^2 + \frac{U}{N} \nonumber (\betadagop{\phi})^2(\betaop{\phi})^2 + G\ \bigg[ (\alphadagop{\phi})^2 + h.c. \bigg] \\
    &+ G\bigg[ (\betadagop{\phi})^2+ h.c. \bigg]  +\frac{\lambda}{N} \bigg[ \alphadagop{\phi}\betaop{\phi} + h.c.\bigg]. \label{effective hamiltonian}
\end{align}
 $\zeta = i\sqrt{NG/U}$ is the same as was defined in section \ref{DFS}. We notice that compared to the uncoupled chain from Zapletal \textit{et al.}~\cite{zapletal_stabilization_2022}, there is now an additional one-photon loss on both subsystems present in the effective master equation. This is due to the coupled model breaking the individual parity symmetry of each chain,  reducing it to total parity through parity switching between chains. This parity exchange between the chains is precisely the mechanism that will later allow us to generate entangling dynamics between the two multi-mode cat qubits. The dissipation rates of two-photon and one-photon loss, respectively, are
\begin{align}
    \Gamma &= 4\frac{U^2}{N^2}\sum_{k\neq \phi}\frac{1}{\gamma_k} =
    \frac{4U^2d}{\gamma N^2} \label{gamma} \\
    \chi &= 4\frac{\lambda^2}{N^2} \sum_{k\neq\phi} \frac{1}{\gamma_k}=\frac{4\lambda^2d}{\gamma N^2}. \label{chi}
\end{align}
Using $1-\cos{x}=2\sin^2{(x/2)}$ and the condition $k\neq \phi$ combined with the restrictions on allowed momenta set by periodic boundary conditions, the sum can be rewritten
\begin{equation}
    \gamma\sum_{k\neq\phi}\frac{1}{\gamma_k}=\frac{1}{4}\sum^{N-1}_{m=1}\frac{1}{\sin^2{(\pi m/N)}}.
\end{equation}
taking $z_m = e^{2\pi i m/N}$ as the non-trivial $N$-th roots of unity, the term inside the sum can be rewritten as $4\abs{1-z_m}^{-2}$. This can again be rewritten into two terms of analytically summable finite series. The result of this calculation is
\begin{equation}
    d = \gamma \sum_{k\neq\phi}\frac{1}{\gamma_k} = \frac{N^2-1}{12}, \label{d definition}
\end{equation}
which is the $d$ in Eqs. \eqref{gamma} and \eqref{chi}. The details of the derivation of the Zeno-limit, including how we obtain $\Gamma$ and $\chi$ are left to Appendix \ref{Appendix Zeno}.
With the system described by Eqs. \eqref{Final lindblad eq} and \eqref{effective hamiltonian}, we can now identify the entangling operation generated by the coupling. Because of the cat-state stabilization we can obtain an effective Hamiltonian by projecting the beamsplitter term in Eq. \eqref{effective hamiltonian} onto the logical subspace. This gives an effective interaction $\propto X \otimes X$ on the two cat qubits~\cite{mirrahimi_dynamically_2014}. Turning on the coupling for a time $t$ implements an $XX$ rotation

\begin{equation}
XX_\theta= e^{-i\frac{\theta}{2} X \otimes X},
\end{equation} 
with the rotation angle $\theta$ depending on the interaction time. Setting $\theta=\pi/2$ realizes the maximally entangling  $XX_{\pi/2}$ rotation. Together with the arbitrary one-qubit rotations from section \ref{Logical subspace}, this entangling gate completes a universal gate set~\cite{bremner_practical_2002}.

\section{ Results}

\subsection{Dynamics}

\begin{figure}
    \centering
    \includegraphics[width=0.9\linewidth]{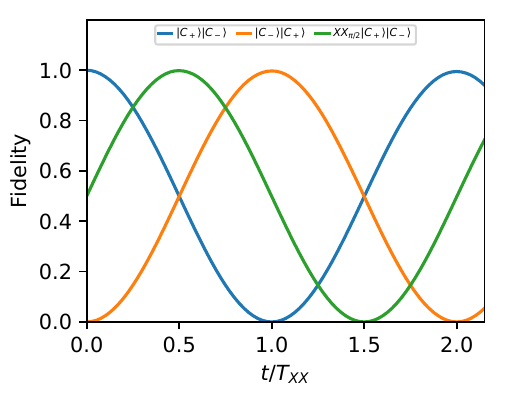}
    \caption{Dynamics of the fidelity of the instantaneous state $\rho_{\phi\phi}(t)$ when $\ket{\psi_0} = \ket{C_+}\ket{C_-}$ against $\ket{\psi_0}$ (blue), $\ket{C_-}\ket{C_+}$ (orange) and $XX_{\pi/2}\ket{\psi_0}$ (green). $U/G = 1$, $\gamma / U=200$ and $\lambda / U=0.1$. $N = 9$ for all simulations. The time parameter is normalized by $T_{XX} = \frac{\pi U}{4\lambda G}$, the period for $XX(\pi)$.}
    \label{fig:Dynamics:fidelity}
\end{figure}

\begin{figure}
    \centering
    \includegraphics[width=0.9\linewidth]{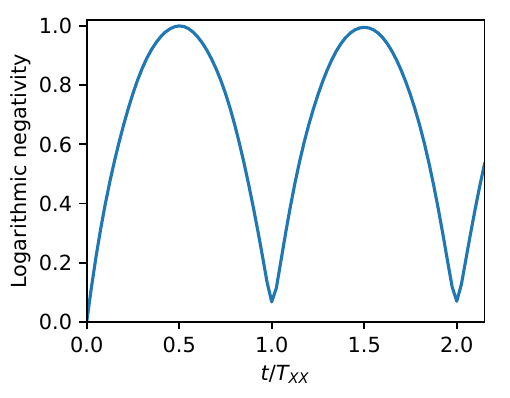}
    \caption{Logarithmic negativity as an entanglement measure of the instantaneous state $\rho_{\phi\phi}(t)$. Parameters are the same as in Fig. \ref{fig:Dynamics:fidelity}.}
    \label{fig:Dynamics:Entropy}
\end{figure}
\begin{figure}[h!]
\captionsetup[subfigure]{labelformat=empty, skip=-10pt}
    \begin{subfigure}{0.5\textwidth}
        \includegraphics[width=0.8\textwidth]{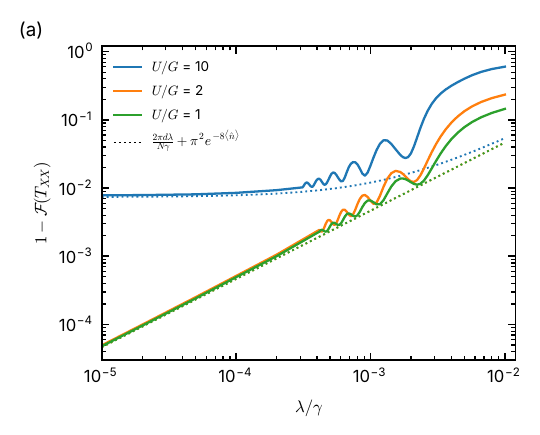}
        \caption{}
    \label{fig:fixed_gamma_200}
    \end{subfigure}
    \begin{subfigure}{0.5\textwidth}
        \includegraphics[width=0.8\textwidth]{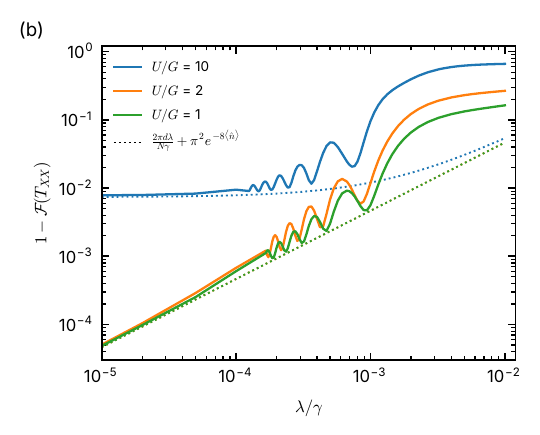}
        \caption{}
    \label{fig:fixed_gamma_500}
    \end{subfigure}
    \begin{subfigure}{0.5\textwidth}
        \includegraphics[width=0.8\textwidth]{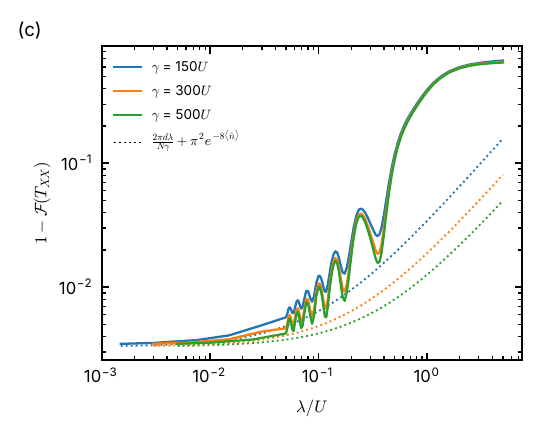}
        \caption{}
    \label{fig:fixed_num}
    \end{subfigure}
    
    \caption{ (a) Resulting infidelities at time $t = T_{XX}$ as a function of $\lambda$ for fixed $U/(N\gamma)$, with $\gamma = 200U$ (a), and $\gamma = 500U$ (b). The blue, orange, and green curves correspond to $U/G=10$, $U/G=2$ and $U/G=1$, respectively. In (c) $\expval{\hat{n}} = NG/U =1$ is fixed, for varying $\gamma=150U$ (blue), $\gamma = 300U$ (orange) and $\gamma = 500U$ (green). With $N=9$, this corresponds to $G=U/N\approx 0.11U$. The dotted lines are a baseline on infidelity induced by one-photon loss and small $\expval{\hat{n}}$ effects, given in Eq. \eqref{1ph and small n bound}. }
    \label{infid_lambda}
\end{figure}

We have simulated the effective two-mode Hamiltonian by solving the master equation Eq. \eqref{Final lindblad eq} numerically using the Python library \textit{QuTiP} ~\cite{johansson2012qutip,qutip5}. In Figs. \ref{fig:Dynamics:fidelity} and \ref{fig:Dynamics:Entropy} we plot the dynamics of the instantaneous state, starting from the product state $\ket{\psi_0}=\ket{C_+}  \ket{C_-}$. Using realistic parameters, we compare the fidelity of the instantaneous state with the initial state, the parity-swapped $XX$ state and the intermediate maximally entangled $XX_{\pi/2}{\ket{\psi_0}}$ state to show how the instantaneous state exhibits Rabi-like oscillations with close to unit fidelity over a full oscillation. We use that the fidelity between two arbitrary states $\rho$ and $\sigma$ is $\mathcal{F(\rho, \sigma)=\Tr\big[\sqrt{\sqrt{\rho}\sigma\sqrt{\rho}}}\big]^2$ to compute this.  We will comment on the  slight decay of fidelity in section \ref{regimes of validity}. We also show the logarithmic negativity, which has previously been used as an entanglement-measure on bipartite Schrödinger cat systems~\cite{liao_generation_2016, vidal_computable_2002, plenio_logarithmic_2005}. This result is in agreement with the simulation of the fidelity, namely that we achieve a logarithmic negativity close to unity for the intermediate state, indicating an entangled state. These simulations show at a glance that the $XX$ operation can be performed with high fidelity, and that we can reach the intermediate entangled state which is a necessary condition for any protocol to achieve universal computation. In~\ref{regimes of validity} (Figs. \ref{infid_lambda} - \ref{fig:Scan over N}) we quantify under what parameter choices this claim is valid and the robustness of the protocol, as Fig.~\ref{fig:Dynamics:fidelity} only shows one configuration as an illustration. In Fig. \ref{fig:Dynamics:Entropy} it is  clear that the logarithmic negativity does not return to zero at multiples of $T_{XX}$, as it would for the expected ideal product state. We simulated this with the induced one-photon loss turned off, and find a residual negativity $\approx 0.07$, meaning this effect is not fully caused by loss. We attribute this to leakage of population outside the DFS of order $\sim \lambda^2$, which adds a non-product component to the ideal product state, with an amplitude, and therefore negativity, which is linear in $\lambda$. This is a coherent effect of the Hamiltonian interaction between two rings, and not an artifact of the Zeno limit. We discuss coherent leakage along with other causes of infidelity further in~\ref{regimes of validity}.  The starting configuration $\ket{\psi_0}=\ket{C_+}\ket{C_-}$ can be achieved with arbitrary fidelity by initializing both decoupled systems in the $\otimes_j\ket{0}_j$ state, and performing a logical $X$-gate on the second subsystem~\cite{zapletal_stabilization_2022}, before turning the coupling on. We therefore only consider the results from this starting point.

\subsection{Contributions to the Infidelity} \label{regimes of validity}
We perform simulations of the gate infidelity over a time $T_{XX}$, by computing $1-\mathcal{F}(\rho(T_{XX}), \ket{C_-}\ket{C_+})$, the infidelity of the instantaneous state against the $XX$-flipped state as a function of the coupling parameter $\lambda$. This analysis gives an estimate for the upper bound of fidelity we can achieve with idealized parameter choices in the proposed protocol, showcasing how the different parameters affect infidelity and robustness of the proposed scheme. The time for the entangling operation can be approximated as $T_{XX}=\pi/\Omega$, where $\Omega = 4\lambda\expval{\hat{n}}/N$ by projecting the interaction term, $\frac{\lambda}{N} \left(\hat{\alpha}_\phi^\dagger \hat{\beta}_\phi+ \text{h.c.} \right)$, down on the tensor product of the cat state manifolds, $\text{span}\left(\ket{C_+}_\phi, \ket{C_-}_\phi\right)$~\cite{mirrahimi_dynamically_2014}. The average excitation number is  $\expval{\hat{n}}=\abs{\zeta}^2=NG/U$, and we get \begin{equation}
    T_{XX} = \frac{\pi U}{4\lambda G}. \label{time one flip}
\end{equation}
 We consider the convention where the period of interest is the time for a single $XX$ flip. In Fig. \ref{infid_lambda} we show the infidelity $1-\mathcal{F}(T_{XX})$ as a function of $\lambda$ for fixed $\gamma = 200U$ (\ref{fig:fixed_gamma_200}) and $\gamma=500U$ (\ref{fig:fixed_gamma_500}) with 3 curves of varying ratio $U/G$. The results show that decreasing $U/G$ \textit{i.e.} increased $\expval{\hat{n}}$ (at fixed $N$) improves fidelity, and for the two smallest ratios, the falloff seems to be linear, while for the largest ratio, the infidelity seems to flatten out to a floor. In Fig. \ref{fig:fixed_num} we also show the infidelity for increasing $\gamma$ with fixed $\expval{\hat{n}}$. Infidelity decreases with increased $\gamma$, consistent with the Zeno-limit. In the following we explain the floor, the linear falloff and the oscillations of the curves in Fig.~\ref{infid_lambda}.

\begin{figure}
    \centering
    \includegraphics[width=0.8\linewidth]{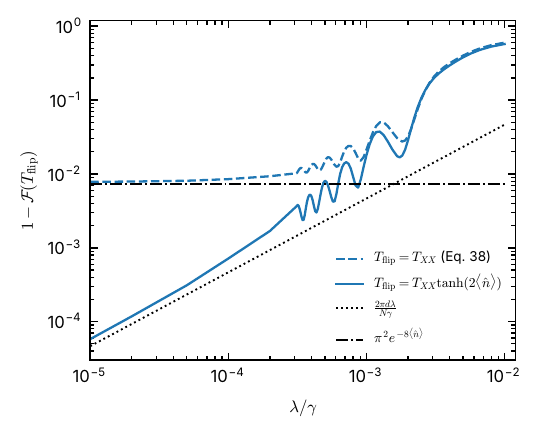}
    \caption{The fidelity over $\lambda/\gamma$ with the adjusted flip time $T_{\text{flip}} = T_{XX}(\expval{\hat{n}})$ accounted for in software (full) and fixed time $T_{\text{flip}}=T_{XX}$ (dashed). The dashed-dotted curve shows the flat small $\expval{\hat{n}}$ floor, while the linear dotted line is the infidelity introduced by one-photon loss.}
    \label{fig:fixed txx}
\end{figure}

Due to the finite size of the cat states, the projection from the beamsplitter interaction to the $XX$-type interaction is an approximation, that is exactly valid in the $\expval{\hat{n}}\to \infty$ limit. The full form of the interaction term, which is derived and shown explicitly in Appendix \ref{Projection XX}, retains a term of the form $YY$. Odd parity input states such as $\ket{C_+}\ket{C_-}$ and $\ket{C_-}\ket{C_+}$ are the worst cases for the fidelity, this gives the limit 
\begin{equation}
    (1-\mathcal{F})_{\expval{\hat{n}}} \approx \pi^2e^{-8\expval{\hat{n}}},\label{infid floor}
\end{equation} 
which is independent of $\lambda$. In Fig. \ref{fig:fixed_num} all three curves plateau at around $4\times10^{-3}$, set by $\pi^2e^{-8\expval{\hat{n}}}\approx 3.3\times10^{-3}$ for $\expval{\hat{n}}=1$. This is about half the plateau of the two blue curves in Figs. \ref{fig:fixed_gamma_200} and \ref{fig:fixed_gamma_500} with $\expval{\hat{n}}=0.9$, showcasing the exponential sensitivity of the floor. Alternatively, by keeping the generalized coupling shown in Eq. \eqref{projection omega}, we can define the $\expval{\hat{n}}$-dependent rate of rotation as
\begin{equation}
    \Omega(\expval{\hat{n}}) = \frac{4\lambda G}{U}\coth{\bigg(\frac{2NG}{U}\bigg)} = \frac{4\lambda G}{U}\coth{(2\expval{\hat{n}})}.
\end{equation}
This implies a shift $T_{XX}$ given as
\begin{equation}
    \frac{T_{XX}(\expval{\hat{n}})}{T_{XX}} = \tanh{\bigg(\frac{2NG}{U}\bigg)} = \tanh{(2\expval{\hat{n}})}. \label{normalized t flip}
\end{equation}
For the specific flip shown here, this eliminates the infidelity floor. This is shown in Fig. \ref{fig:fixed txx} where we recreate the blue curve from Fig. \ref{fig:fixed_gamma_200} with fidelity measured at $T_{XX}(\expval{\hat{n}})$. For a constant value of $T_{XX}$, this term dominates at $\lambda/\gamma\sim 10^{-4}$ , while when correcting for the shifted period, the falloff remains linear and closely follows the curve described by $2\pi d\lambda/(N\gamma)$. This term comes from induced one-photon loss, which we now describe.

\begin{figure}
    \centering
    \includegraphics[width=0.8\linewidth]{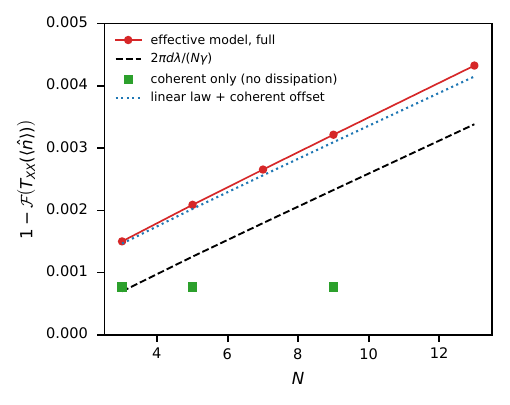}
    \caption{Scan over the values $N=3, 5, 7, 9, 13$ of the infidelity at corrected time $T_{XX}(\expval{\hat{n}})$. $\expval{\hat{n}}=4$, $\lambda=0.1U$ and $\gamma=200U$. $U/G$ is fixed by $N$ and $\expval{\hat{n}}$. The full red curve shows the simulated infidelity of the effective model from equation Eq. \eqref{Final lindblad eq}, the dashed line shows the linear law of one photon loss from Eq. \eqref{linear one ph loss}, the green points show an N-independent offset due to coherent leakage. The dotted line is the sum of the previous two effects.}
    \label{fig:Scan over N}
\end{figure}

The jump-operator $\sqrt{\chi}\alphaop{\phi}$ acts as a logical bitflip on the A subsystem, at the rate
\begin{equation}
    \bra{C_\pm}\chi\alphadagop{\phi}\alphaop{\phi}\ket{C_\pm} = \chi\abs{\zeta}^2r^{\pm2} = \chi\expval{\hat{n}} \big[1 + O(e^{-2\expval{\hat{n}}})\big]
\end{equation}
where
\begin{equation}
    r = \frac{\mathcal{N}_+}{\mathcal{N}_-} \label{r def}
\end{equation}
with $\mathcal{N}_\pm$ defined under Eq. \eqref{cat states def}. Including the B subsystem, the total rate of jumps is therefore $2\chi\expval{\hat{n}}$ to lowest order. Since $X$ on either the A or B subsystem commutes with the logical $XX$ gate, a jump at any time $t\leq T_{XX}$ commutes through the evolution and will leave the resulting state orthogonal to the target state. The total probability of a jump over the time $T_{XX}$ is therefore $p = 2\chi \expval{\hat{n}}T_{XX}$. By inserting $\chi$ from Eq. \eqref{chi} and $T_{XX}$ from Eq. \eqref{time one flip} this affects infidelity as
\begin{equation}
    (1-\mathcal{F})_\lambda = \frac{2\pi d\lambda}{N\gamma}. \label{linear one ph loss}
\end{equation}
This shows the tradeoff between induced dissipation that grows as $\chi \propto \lambda^2$ and gate time, which shrinks as $T_{XX} \propto 1/\lambda$ such that the error grows only linearly in $\lambda$. Combining the two aforementioned effects gives
\begin{equation}
    1-\mathcal{F} \approx \frac{2\pi d \lambda}{N\gamma} + \pi^2e^{-8\expval{\hat{n}}}, \label{1ph and small n bound}
\end{equation}
and is used as a reference curve serving as a baseline in Fig. \ref{infid_lambda}. 
We finally consider the effects of the beamsplitter interaction from Eq. \eqref{H BS}, which does \textit{not} confine states to the DFS. Through this, the states $\ket{C_\pm}$ are coupled to states outside the DFS at an energy gap of approximately $4G$, resulting in a leaked population that oscillates coherently over time. To lowest order in $\lambda/U$ the average leaked population is
\begin{equation}
    \overline{p} = \frac{\lambda^2}{4\expval{\hat{n}}U^2}. \label{coherent leakage}
\end{equation}
Leakage affects the gate in two ways. The population outside the DFS at the gate time is a coherent loss of fidelity, which we call $p_{coh}$. Secondly, the leaked population is no longer dark, and is susceptible to loss through $\hat{Z}_\phi = \hat{z}_\phi^2-\zeta^2 \text{, } \hat{z}_\phi \in \{\alphaop{\phi}, \betaop{\phi}\}$ from Eq. \eqref{Final lindblad eq}. On the leaked states, $\expval{\hat{Z}^\dagger_\phi\hat{Z}_\phi} \simeq 4\expval{\hat{n}}\overline{p} = \lambda^2/U^2$. For the full gate time this gives
\begin{equation}
    (1-\mathcal{F})_\Gamma = \xi\Gamma T_{XX} \frac{\lambda^2}{U^2} = \frac{\xi\pi d\lambda}{N\gamma\expval{\hat{n}}}. \label{infid 2ph}
\end{equation}
It is linear in $\lambda$ and follows the same $d/(N\gamma)$ dependence as the one-photon loss, but is suppressed by $\xi/(2\expval{\hat{n}})$. $\xi \leq 1$ is a factor that accounts for jumps returning leaked states to the DFS as either target-like or initial-like states, so not every jump causes infidelity. It is not obtained in closed form, but numerical analysis suggests $\xi \approx 0.3-0.75$.
The leaked population oscillates between $0$ and $2\overline{p}\propto \lambda^2/(\expval{\hat{n}}U^2)$ at a frequency close to $4G$. At $t=T_{XX}$ the accumulated phase therefore scales as $4GT_{XX} \propto U/\lambda$. As $\lambda$ (and $T_{XX}$) is varied this runs through several multiples of $2\pi$, which is the cause of the harmonic ripples observed in Figs. \ref{infid_lambda} and \ref{fig:fixed txx}. The ripples are visible only at intermediate $\lambda$ because he leakage amplitude grows as $\lambda^2$, while the baseline Eq. \eqref{1ph and small n bound} is linear, so for small $\lambda$ the ripples are far below the baseline. They are visible when the two are comparable, at $\lambda\approx 0.1U-0.5U$ using the parameters of Fig. \ref{infid_lambda}. Equation \eqref{1ph and small n bound} serves as a baseline to which coherent leakage and Eq. \eqref{infid 2ph} add.
Figure \ref{fig:Scan over N} separates these contributions in a scan over N, at fixed $\expval{\hat{n}}=4$, $\lambda=0.1U$ and $\gamma=200U$. The infidelity at $T_{XX}$ grows linearly in $N$ with the contribution from Eq. $\eqref{linear one ph loss}$, on top of an $N$-independent (at fixed $\expval{\hat{n}})$ offset from coherent leakage. Adding the second linear in $N$ term from Eq. \eqref{infid 2ph} with $\xi\approx0.4-0.45$ closes the gap with the simulated infidelity. Keeping all effects, a bound on infidelity can be given as

\begin{equation}
    1-\mathcal{F} \approx \frac{2\pi d \lambda}{N\gamma} + \frac{\xi \pi d\lambda}{N\gamma\expval{\hat{n}}} + p_{coh}(\lambda/U, \expval{\hat{n}}) + \pi^2e^{-8\expval{\hat{n}}}, \label{infid budget}
\end{equation}
where a closed form of the coherent leakage is not derived.
In Appendix \ref{Full dynamics} we compare the effective model to the full genuine multi-mode model of Eq. \eqref{two mode lindblad}, which accounts for an additional inaccuracy due to the Zeno-approximation and adds a final term $O(\gamma^{-2})$ to Eq.~\eqref{infid budget}.

\subsection{Intrinsic Single-Photon Loss\label{subsec:intrinsic-loss}}

In most current circuit-QED implementations, single-photon loss at each oscillator is the dominant intrinsic decoherence channel~\cite{zapletal_stabilization_2022}. We model it on both rings by
\begin{equation}
\mathcal{L}_\kappa\rho
= \kappa\sum_{j=1}^N\big(D[\aop{j}]+D[\bop{j}]\big)\rho
= \kappa\sum_{k}\big(D[\alphaop{k}]+D[\betaop{k}]\big)\rho,
\label{intrinsic loss}
\end{equation}
where the second equality follows because the Fourier transform Eq. \eqref{fourier transform op} is unitary. Intrinsic loss is therefore uniform in the normal-mode basis. It shifts every dissipative rate as $\gamma_k\to\gamma_k+\kappa$. For the bright modes $k\neq\phi$ this only strengthens the dissipative protection, while for the protected modes it opens the single-photon channels $D[\alphaop{\phi}]$ and $D[\betaop{\phi}]$ acting \emph{inside} the cat manifold. Using  $\alphaop{\phi}\ket{C_\pm}=\zeta(\tanh\abs{\zeta}^2)^{\pm1/2}\ket{C_\mp}$ (and the analogous relation for $\betaop{\phi}$), each such channel acts as a logical parity flip at rate $\kappa\abs{\zeta}^2(\tanh\abs{\zeta}^2)^{\pm1}\approx$ $\kappa\abs{\zeta}^2=N\kappa G/U$. During the $XX$ it simply adds to the induced rate Eq. \eqref{chi}, $\chi\to\chi+\kappa$, and contributes $2\kappa\abs{\zeta}^2T_{XX}=\pi N\kappa/(2\lambda)$ to the gate infidelity. For the input $\ket{\psi_0}=\ket{C_+}\ket{C_-}$ each jump costs the full fidelity, since a single jump changes total photon parity and therefore produces a state orthogonal to the target. The gate therefore requires $\kappa\ll\lambda/N$.

The induced loss acts in the same way. Adding the intrinsic loss to the reference curve of Eq. \eqref{1ph and small n bound} gives
\begin{equation}
\begin{split}
1-\mathcal{F}&\simeq\frac{\pi N(\chi+\kappa)}{2\lambda}+\pi^2e^{-8\expval{\hat{n}}}\\
&=\frac{2\pi d}{N}\,\frac{\lambda}{\gamma}+\frac{\pi N\kappa}{2\lambda}+\pi^2e^{-8\expval{\hat{n}}},
\end{split}
\label{loss budget}
\end{equation}
with $d$ defined in Eq. \eqref{d definition}. Without intrinsic loss, the above estimate grows monotonically with $\lambda$. With it, the minimum of this estimate is obtained by setting $\chi=\kappa$ in Eq.~\eqref{chi}. This gives
\begin{equation}
\lambda_{\mathrm{opt}}=N\sqrt{\frac{\kappa\gamma}{4d}},\qquad
1-\mathcal{F}_{\min}\simeq2\pi\sqrt{\frac{d\kappa}{\gamma}}+\pi^2e^{-8\expval{\hat{n}}},
\label{loss optimum}
\end{equation}
which is the trade-off between gate speed and induced dissipation. For $N=9$, $G=U$, $\gamma=200U$ and $\kappa=10^{-4}U$ we get $\lambda_{\mathrm{opt}}\approx0.25U$ and $1-\mathcal{F}_{\min}\approx1.1\times10^{-2}$. A direct simulation of the effective model gives $1.2\times10^{-2}$, with corrections from the leakage terms of Eq.~\eqref{infid budget}. At fixed $\kappa/\gamma$ the minimal error grows as $\sqrt{N^2-1}$. It is nearly independent of $N$ if $\gamma$ grows as $N^2$, which keeps the smallest bright rate $4\gamma\sin^2(\pi/N)$ fixed.

For preparation and idling, the key quantity is the ratio of the parity-flip rate to the dissipative gap $\Gamma_d$ protecting the decoherence-free subspace~\cite
{zapletal_stabilization_2022}. The budget Eq. \eqref{loss budget} does not depend on $\Gamma_d$. A photon loss moves the state from one cat to the other, and the parity-conserving confinement cannot undo this. In the Kerr model (K) considered here, the gap is set by the Zeno-induced two-photon dissipation,
\begin{equation}
	\Gamma_d^{(K)}\;\approx\;2\Gamma\abs{\zeta}^2
	=\frac{8d}{N}\,\frac{UG}{\gamma}
	\qquad[\text{cf. Eq.~\eqref{gamma}}],
	\label{kerr gap}
\end{equation}
so that $\kappa\abs{\zeta}^2/\Gamma_d^{(K)}\approx\tfrac{3}{2}\kappa\gamma/U^2$. The same large $\gamma$ required for the Zeno reduction of Sec.~\ref{zeno limit coupled chains} suppresses the gap and amplifies the relative impact of intrinsic loss, which on the single-ring level
already limits the preparation fidelity at moderate $\kappa/U$~\cite[Sec.~VII]{zapletal_stabilization_2022}. Within model (K) this tension can be reduced. The Zeno reduction only needs the smallest bright rate $4\gamma\sin^2(\pi/N)$ to exceed $U$, $G$ and $\lambda$, so $\gamma$ should be kept at this edge. The gate prefers a larger $\gamma$ by Eq. \eqref{loss optimum}, and a $\gamma$ that is tunable in time would reconcile the two. The parity flips themselves commute with the gate and are left to an outer code, as in bias-preserving architectures~\cite{puri_bias-preserving_2020,chamberland2022b}. In the alternative model (L) of~\cite[Sec.~VI]{zapletal_stabilization_2022}, where the on-site Kerr nonlinearity is replaced by engineered local two-photon loss at rate $\eta$, the gap $\Gamma_d^{(L)}\approx2G$ is independent of
$\gamma$, the ratio $\kappa\abs{\zeta}^2/\Gamma_d^{(L)}=$  $\kappa N/(2\eta)$ carries no Zeno penalty, and correspondingly higher preparation fidelities are obtained under the same $\kappa$~\cite[Sec.~VII]{zapletal_stabilization_2022}.

For the two-ring gate the picture is unchanged. The effective master equation Eq. \eqref{Final lindblad eq} acquires the additional dissipators $\kappa\big(D[\alphaop{\phi}]+D[\betaop{\phi}]\big)$ on the protected modes, and the $XX$ remains reliable as long as $\kappa\ll\lambda/N$.  Model (L) reduces in the same way. Its local two-photon loss projects at first order onto $(2\eta/N)D[\hat{z}_\phi^2-\tilde\zeta^2]$ on each ring, with $\tilde\zeta^2=-iNG/\eta$~\cite[Sec.~VI]{zapletal_stabilization_2022}. The effective equation is then Eq. \eqref{Final lindblad eq} with $\hat H_\phi$ reduced to the beam-splitter term, the $\Gamma$ term replaced by this dissipator and the same $\chi$. The leading single-photon-loss contribution is the same in both models. Equations~\eqref{loss budget} and \eqref{loss optimum} therefore give a common loss-only estimate. Leakage of order $\lambda^2$ of Sec.~\ref{regimes of validity} can shift the total error and its optimum. In model (K) the energy gap $4G$ above the cats keeps it small and oscillating. In model (L) the dissipator pumps it back at a rate of order $2G$. We therefore do not claim an advantage of model (L) for the two-qubit gate. Its advantage is the faster preparation under intrinsic loss, which also frees $\gamma$ to be raised to lower Eq. \eqref{loss optimum}.


\subsection{Robustness Against Imperfections\label{subsec:robustness}}

In a realistic implementation the engineered rates $\gamma$ and bond phases $\phi$ in Eq.~\eqref{dissipator} are not perfectly homogeneous. We replace them on each ring $R\in\{A,B\}$ by site-dependent rates $\gamma_j^{R}\geq 0$ and bond phases $\phi_j^{R}$, with harmonic mean rate $\overline\gamma^{R}\equiv$  $\big(\frac1N\sum_{j=1}^N1/\gamma_j^{R}\big)^{-1}$, giving an inhomogeneous dissipator whose normal modes are no longer plane waves. Inhomogeneous data  can prevent a perfect dark mode through a global obstruction in the \emph{loop phase} (the overall phase in~\cite
{zapletal_stabilization_2022})
\begin{equation}
\Phi^{R}=\sum_{j=1}^N\phi_j^{R}\pmod{2\pi},
\label{loop phase}
\end{equation}
which must vanish for a perfect dark mode to exist~\cite
{zapletal_stabilization_2022}. A dark mode drops out of every jump operator, so its site amplitudes must satisfy $u_j=e^{i\phi_j^R}u_{j+1}$. Going around the ring shows that this is possible exactly when $\Phi^R=0$, for any positive rates. Generic disorder spoils this condition, and the selected normal mode acquires a small residual single-photon loss $\widetilde\gamma_m^{R}>0$ that adds to the intrinsic loss Eq.~\eqref{intrinsic loss} \emph{inside} the cat manifold. For equal rates and a uniform phase mismatch $\delta\phi^{R}$ on top of the design value, with $\abs{\delta\phi^R}<\pi/N$,
\begin{equation}
\widetilde\gamma_m^{R}
= 2\overline\gamma^{R}\big(1-\cos\delta\phi^{R}\big)
\;\simeq\;\overline\gamma^{R}\big(\delta\phi^{R}\big)^2.
\label{residual uniform}
\end{equation}
For bond-by-bond phase disorder the obstruction is global rather than local. Choosing the mode profile that optimally absorbs the bond phases leaves a residual loss set only by the loop defect, $\widetilde\gamma_m^{R}\approx\overline\gamma^{R}\,(\Phi^R)^2/N^2$, where now $\Phi^R=\sum_j\delta\phi_j^{R}$ (mod $2\pi$). The optimal profile spreads the loop phase over the bonds, with a slip $\Phi^R\overline\gamma^R/(N\gamma_j^R)$ on bond $j$. This gives the harmonic mean. Independent bond phases of variance $\sigma^2$ give $\langle\widetilde\gamma_m^{R}\rangle\approx\overline\gamma^{R}\sigma^2/N$. This residual is quadratic in the disorder strength. The flip rate $\widetilde\gamma_m^{R}\abs{\zeta}^2$ does not decrease with $N$, since $\abs{\zeta}^2=NG/U$. Disorder in the rates $\gamma_j^{R}$ alone leaves the selected mode exactly dark, since the dark-mode condition involves only the phases.  Single-photon amplification can instead arise when phase disorder changes the dark-mode profile without a corresponding adjustment of the pumps~\cite[App.~H]{zapletal_stabilization_2022}.

A missing bond turns the ring into an open chain without a loop condition. A dark mode then exists for any bond phases, but for equal rates the smallest bright rate drops from $4\gamma\sin^2(\pi/N)$ to $4\gamma\sin^2(\pi/2N)$. The rings also need not share the same $\phi$. For design phases $\phi^A$ and $\phi^B$ the coupling~\eqref{H BS} acquires the phase $\theta_{BS}=l\phi^A-l'\phi^B$ on the protected modes, which reduces the projected interaction by $\cos\theta_{BS}$. A static phase on $\lambda$ restores it.
	
Inhomogeneous Kerr rates $U_j$ and pumps $G_j$ enter the Hamiltonian, not the dissipator, and leave the dark mode unchanged. Up to a constant, the local Hamiltonian on site $j$ reads
\begin{equation}
U_j\big[(\adagop{j})^2-\zeta_j^{*2}\big]\big[\aop{j}^2-\zeta_j^2\big]+\big[\delta_j(\adagop{j})^2+h.c.\big],
\label{mismatch}
\end{equation}
with $\zeta_j$ the site amplitudes of Sec.~\ref{DFS} and the mismatch $\delta_j=G_j+U_j\zeta_j^2$. If all $\delta_j$ vanish, Eqs. \eqref{schrodinger eq} and \eqref{criteria operators} hold exactly for arbitrary $U_j>0$, without the Zeno limit. A measured spread of the $U_j$ can thus be compensated by the pumps, $\abs{G_j}=U_j\abs{\zeta}^2/N$. An uncompensated mismatch only renormalizes $\zeta$ at first Zeno order. For homogeneous Kerr coefficients, pump mismatch gives a second-order heating term $\mu D[\alphadagop{\phi}]$ with $\mu=4\sum_{q\neq\phi}\abs{M_q}^2/\gamma_q$ and $M_q=(2/N)\sum_j\delta_je^{i(\phi+q)j}$. Independent, zero-mean mismatches with $\langle|\delta_j|^2\rangle=\sigma_\delta^2G^2$ give $\langle\mu\rangle=16d\sigma_\delta^2G^2/(N\gamma)$. Heating flips parity and can cause leakage; treating $\mu$ as an additional $\kappa$ in Eq.~\eqref{loss budget} is a large-cat approximation. For $N=9$, $G=U$, $\gamma=200U$ and $\sigma_\delta=0.05$ we get $\mu\approx1.5\times10^{-4}U$. Phase noise of only $\sigma=0.01$ gives $\langle\widetilde\gamma_m^{R}\rangle\approx2.2\times10^{-3}U$, since phase errors are not Zeno-suppressed. A spread of the oscillator frequencies with variance $\sigma_\omega^2$ instead adds a single-photon loss of rate $\expval{\chi_\omega}=4d\sigma_\omega^2/(N\gamma)$ on the protected mode.

 To leading order, independent residual single-photon losses on the two rings contribute additively to the error budget. The effective master Eq.~\eqref{Final lindblad eq} acquires the additional loss $\widetilde\gamma_m^{A}D[\alphaop{\phi}]+ \widetilde\gamma_m^{B}D[\betaop{\phi}]$ on the protected modes, and the Zeno reduction of Sec.~\ref{zeno limit coupled chains} remains valid as long as $\min_{k\neq m}\widetilde\gamma_k^{R}\gg U,G,\lambda$ on each ring. In Eq. \eqref{loss budget} these rates enter as $\kappa\to\kappa+(\widetilde\gamma_m^A+\widetilde\gamma_m^B)/2$, and independent $\chi_\omega$ contributions likewise. Bond-phase errors dominate this budget. Their loss is not Zeno-suppressed and exceeds $\mu$ for $\sigma=\sigma_\delta$ by the factor $\gamma^2/(16dG^2)$, about $375$ for the parameters above.

The Kerr stabilization model is again the more fragile choice for preparation. The disorder-induced flip rate $\widetilde\gamma_m^{R}\abs{\zeta}^2$ must be compared with the Zeno-suppressed gap $\Gamma_d^{(K)}\propto UG/\gamma$ of Eq.~\eqref{kerr gap}, giving a ratio $\propto\widetilde\gamma_m^{R}\gamma/U^2$ that grows with the very $\gamma$ the scheme relies on. The two-photon-loss model has the same residual rate $\widetilde\gamma_m^{R}$ but the larger, $\gamma$-independent gap $\Gamma_d^{(L)}\approx2G$, and tolerates substantially more disorder. For that model, moderate disorder  ($25\%$ in the rates and $0.26$\,rad in the phases) still allows cat-state preparation with high fidelity on average for $N=3$~\cite[Sec.~VIII]{zapletal_stabilization_2022}.  For the gate, the disorder-induced rates enter Eq. \eqref{loss budget} in the same way in both models. Conversely, the same sensitivity of the dark mode to the engineered rates and phases suggests using deliberately inhomogeneous $\gamma_j$ and $\phi_j$ as a resource for state engineering.

\section{Conclusion and Outlook}
We have  extended an existing idea for fault tolerant multi-mode Schrödinger cat quantum memory to a full scheme for quantum operations as a necessary requirement on the path to performing universal quantum computing. Upgrading the existing scheme does not introduce a big engineering challenge past the largest barrier for the setup, which is the engineered non-local dissipation. This engineering challenge is the building block of the multi-mode cats  the whole protocol is based on, and achieving this at all, and certainly with a high dissipation rate without introducing additional detrimental effects, is the limiting factor. The beauty of our proposed protocol is the relative simplicity of the remaining challenges in achieving full quantum control of the qubits. We project the full model on a two-mode model, allowing us to build on previous results. We confirm numerically and analytically that the reduction is valid under our parameter choices and what parameters control the accuracy, while showing that the reduction does not alter the physics of performing the proposed quantum gates for the values ($N=3$, $XX$-gate and $N=5$, $Z$-gate) used to compare. The case for $N=9$ is only simulated for the effective model, and the conclusion rests on the estimate that gate error grows as $N/\gamma$, meaning the rate of non-local dissipation must be increased for larger $N$, which is limited by bond-phase disorder (Sec.~\ref{subsec:robustness}).  Section~\ref{subsec:robustness} treats non-identical oscillators to leading order. A full simulation with disorder is left for future work. For future physical implementations, an aim should be to also establish another, more hardware efficient, physical mechanism than the non-local dissipation to stabilize the multi-mode cats, and extend that protocol to be universal. No algorithms were demonstrated in the current work, and neither was a systematic comparison of resource overhead presented. Future work should take both into consideration.

\begin{acknowledgments}
JB and TO are partially funded by The Research Council of Norway, projects 324944 and 345433, respectively, and project 361617 jointly. The authors give thanks to the anonymous referees, whose constructive feedback helped improve this work.

\end{acknowledgments}

\appendix
\section{One qubit rotations in the DFS} \label{appendix rotations}

\subsection{$X(\alpha)$}
A single photon drive applied to an arbitrary labeled oscillator $j'$ through the driving Hamiltonian  
\begin{equation}
    H_\eta = \eta \left( e^{i j' \phi} \hat{a}_{j'} + e^{-i j' \phi} \hat{a}_{j'}^\dagger \right)
\end{equation}
will implement arbitrary rotations around the $X$-axis in the decoherence free subspace. The Zeno-approximation preserves this term, which manifests as an effective one-photon drive on the protected $\phi$-mode, all other terms the same. The effective Hamiltonian becomes
\begin{equation}
    \hat{H}_X = \frac{U}{N}((\alphadagop{\phi})^2-\zeta^{*2})(\alphaop{\phi}^2-\zeta^{2}) +\frac{\eta}{\sqrt{N}}(\alphaop{\phi}+\alphadagop{\phi}),
\end{equation}
with two-photon dissipation still in place as a dissipative stabilizer. From~\cite{mirrahimi_dynamically_2014}, this gives rise to the rotation
\begin{align}
    \begin{split}
        \hat{R}(\alpha) = \cos{\alpha}\big(\ket{C_+}\bra{C_+} + \ket{C_-}\bra{C_-}\big) \\
        + i\sin{\alpha}\big(\ket{C_+}\bra{C_-} + \ket{C_-}\bra{C_+}\big),
    \end{split}
\end{align}
which allows for arbitrary rotation around the $X$-axis.

\subsection{ $Z(\pi/2)$}
To implement the second rotation, a $\pi/2$ rotation around the $Z$-axis, we still consider the chain of resonators coupled dissipatively to each other and with periodic boundary conditions.  The setup with two-photon stabilization renders the states $\ket{\pm} \propto (\ket{C_+}\pm\ket{C_-})$ highly stable. This contradicts the aim of a rotation around the $Z$-axis. We stay in the frame chosen in Eq. \eqref{Hamiltonian real space}, with the pump amplitude at $G=G_{prep}$ and stabilize a cat. We then set $G=0$ for a time $t=T_Z$, while the dissipator Eq.  \eqref{dissipator} stays on. The Hamiltonian during this operation is
\begin{equation}
          \hat{H} =  U \sum_j (\adagop{j})^2(\aop{j})^2 .\label{full hamiltonian z}
\end{equation}
Projecting on the dark mode, the effective single mode description of the system in the strong dissipation limit is
\begin{equation}
    \dot{\rho}_{\phi\phi}(t) = -i\comm{\hat{H}_\phi}{\rho_{\phi\phi}(t)} + \Gamma D[\alphaop{\phi}^2]\rho_{\phi\phi}(t), \label{master eq z}
\end{equation}
with the effective Hamiltonian
\begin{equation}
    \hat{H}_\phi = \frac{U}{N}(\alphadagop{\phi}\alphaop{\phi})^2  -\frac{U}{N}\alphadagop{\phi}\alphaop{\phi}. \label{Z effective Hamiltonian}
\end{equation}
The non-local dissipator enters only through $\Gamma$ at second order in the Zeno-approximation, with the jump-operator shifted because $\zeta =0$ when $G=0$. The quadratic term was shown by Yurke  ~\cite{yurke_generating_1986}  to generate Schrödinger cat states with an arbitrary phase difference. In particular, with the self-Kerr phase
\begin{equation}
    e^{-i(U/N)\hat{n}^2T_Z} = e^{-i(\pi/2)\hat{n}^2} 
\end{equation}
for $T_Z=N\pi/2U$, $\ket{C_\pm}$ are both eigenstates with eigenvalues $+1$ and $-i$ respectively. Under this rotation, the states $\ket{\pm}_L=(1/\sqrt{2})\big(\ket{C_+} \pm \ket{C_-}\big)$ experience the transformations
\begin{equation}
    \ket{\pm} \xrightarrow{t=T_Z} \frac{1}{\sqrt{2}}\big(\ket{C_+} \mp i\ket{C_-}\big),
\end{equation}
showing a rotation by angle $\pi/2$ around the $Z$-axis for $\ket{+}$. The linear term is the induced rotation in phase-spaces. At $t=T_Z$, the cats now lie on the axis orthogonal to the stabilized axis, and when turning the pump back on this must be accounted for by fixing the phase of the pump to stabilize along the new axis.

\begin{figure}[h!]
\centering

    \includegraphics[width=0.45\textwidth]{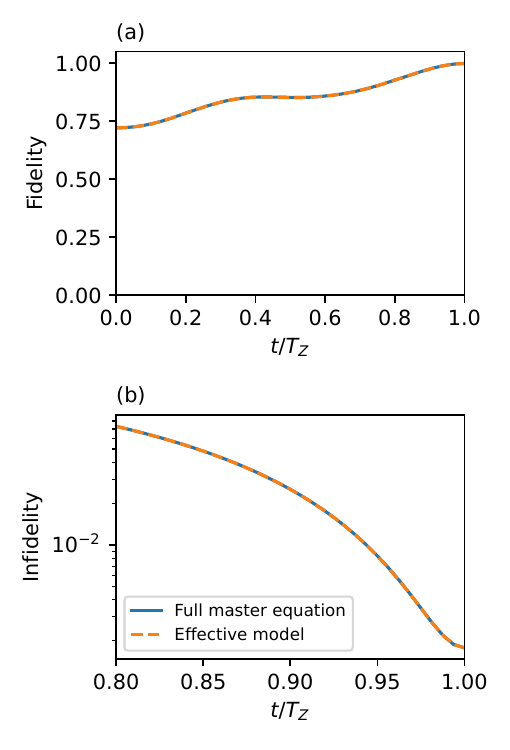}
    
    \caption{(a) Dynamics of the fidelity of the instantaneous state $\rho(t)$ during a $Z_{\pi/2}$-gate from the initial state $\ket{+} = (1/\sqrt{2})\big(\ket{C_+}+\ket{C_-}\big)$ with $\gamma=200U$, prepared with $G = G_{prep}/U = 0.1$ for $N=5$ against the target state $(1/\sqrt{2})\big(\ket{C_+}-i\ket{C_-}\big)$ (with $\ket{C_\pm}$ defined along the rotated axis) for the full model Eq. \eqref{full hamiltonian z} with dissipation from Eq. \eqref{dissipator} (blue) and
     the effective Zeno-limit master equation of Eq. \eqref{master eq z} (dashed orange). (b) Zoomed infidelity approaching $t=T_{Z}$.}
    \label{fig_rot both}
\end{figure}

 To check that the Zeno-reduction of the $Z$-rotation holds, we have simulated the full master equation with Hamiltonian from Eq. \eqref{full hamiltonian z} transformed to the Fourier basis for $G=0$ and $N=5$ and the non-local dissipator of Eq. \eqref{dissipator k space} (also with $G=0$). We consider the ideal state $\ket{+}$ on the dark mode at $\expval{\hat{n}_j}=0.1$, which can be prepared at $G_{prep}/U = 0.1$ and $\gamma=200U$ and compare its fidelity with $Z_{\pi/2}\ket{+} = (1/\sqrt{2})\big(\ket{C_+} - i\ket{C_-}\big)$, where the cat states of the target now lie on the axis rotated by $\pi/2$ due to the linear term in Eq. \eqref{Z effective Hamiltonian}. In the simulations $G$ is turned off instantaneously after preparation. This is compared with the results of simulating the effective model of Eq. \eqref{master eq z} directly. The results are shown in Fig. \ref{fig_rot both} and show that the two models match to about $10^{-5}$ at all times simulated and $2\times10^{-6}$ at $t= T_Z$. Note the choice of parameters leaving $\expval{\hat{n}} = N\expval{\hat{n}_j} = 0.5$ smaller than the examples shown in Fig. \ref{fig:Dynamics:fidelity}. Since the small $\expval{\hat{n}}$-infidelity to the $XX$ gate can be corrected in software, using small $\expval{\hat{n}_j}$ for $Z_{\pi/2}$ does not introduce additional errors for $XX$. The infidelity at time $t=T_Z$ is of order $1.8\times10^{-3}$, much larger than the difference between the full and effective model, and is caused by the two-photon loss when the state leaves the DFS during the operation, $\Gamma T_Z\sim 0.013$ at the chosen parameters, and is consistent within the Zeno-limit.

\subsection{Arbitrary Single Qubit Rotations}

To see that the rotations $X(\alpha)$ and $Z(\pi/2)$ generate arbitrary single qubit rotations, consider that any $U \in SU(2)$ can be generated by three rotations of suitable angles $\alpha, \beta, \gamma$ around two non-collinear axes~\cite{Nielsen_Chuang_2010}
\begin{equation}
    U = X(\alpha)Y(\beta)X(\gamma).
\end{equation}
Using the additional relation
\begin{equation}
    Y(\beta) = Z(\pi/2)^\dagger X(\beta)Z(\pi/2)
\end{equation}
we can rewrite the generating equation as
\begin{equation}
     U = X(\alpha)Z(\pi/2)^\dagger X(\beta)Z(\pi/2) X(\gamma)
\end{equation}
which formally generates $SU(2)$ from the current operations from three arbitrary $X$-rotations, where one is conjugated by $Z(\pi/2)$.
\section{Lindbladian Dynamics of the Full Model}
\label{Full dynamics}
\begin{figure}[!h]
\centering
\captionsetup[subfigure]{labelformat=empty, skip=-10pt}
    \begin{subfigure}{0.5\textwidth}
        \includegraphics[width=0.7\textwidth]{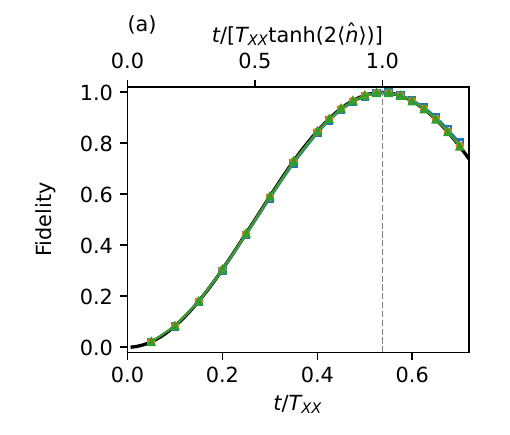}
    
    \caption{}
    \label{fig:full}
    \end{subfigure}
    \begin{subfigure}{0.5\textwidth}
        \includegraphics[width=0.75\textwidth]{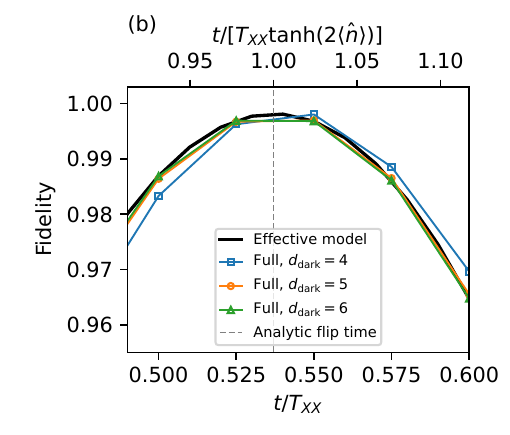}
    
    \caption{}
    \label{fig:full zoomed}
    \end{subfigure}
    \begin{subfigure}{0.5\textwidth}
        \includegraphics[width=0.75\textwidth]{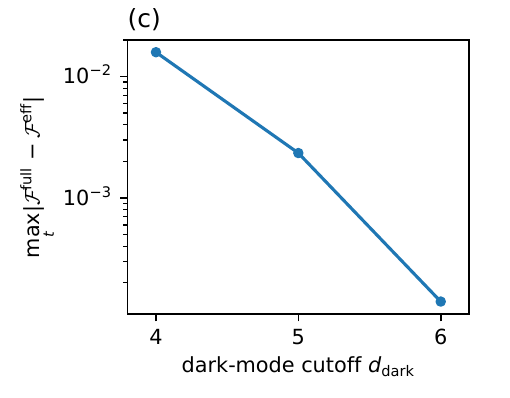}
    
    \caption{}
    \label{fig:Fock_convergence}
    \end{subfigure}
    \caption{(a) The fidelity of the instantaneous state $\rho_{\phi\phi}(t)$ from the effective model (black) and the full model with Fock space cutoff $d_{dark} =4,5,6$ for the blue, orange and green curves respectively, against the target state $\ket{\psi}=\ket{C_-}\ket{C_+}$. On the lower $t$-axis, time $t$ is normalized by the approximate $T_{XX}$ from Eq. \eqref{time one flip}, while the upper axis is normalized by the small $\expval{\hat{n}}$ fixed value of $T_{XX}(\expval{\hat{n}})$ from Eq. ~\eqref{normalized t flip}. (b) Shows the same curves zoomed in around the corrected $T_{XX}(\expval{\hat{n}})$, with the grey dashed line indicating the numerically calculated value. (c) The maximum mismatch over a full flip as a function of Fock-space cutoff of the dark mode $d_{dark}$. In all subfigures $N=3$, $\gamma = 200U$,$\expval{\hat{n}} = NG/U = 0.3$ and $\lambda=0.1U$.}
    \label{fig: full model simulations}
\end{figure}

We compare results of the full multi-mode Lindblad master equation Eq. \eqref{two mode lindblad} with the Zeno-reduced effective two-mode Lindblad equation Eq. \eqref{Final lindblad eq} to analyze the robustness of the Zeno limit in the current case. We simulate for the case $N=3$ on both chains, and we perform the simulation with a Fock-space cutoff $d_{bright} = 2$ on the bright modes as these modes will be heavily suppressed by the dissipation. The simulations are performed at low average excitation number. As we argued in the Sec. \ref{regimes of validity}, the small $\expval{\hat{n}}$ infidelity can be corrected by shifting time $T_{XX}$. We therefore implement this here. The fidelity over one flip is shown in Fig. \ref{fig: full model simulations} for the case $N=3$ and $\gamma = 200U$. We use $\expval{\hat{n}} = NG/U = 0.3$ and $\lambda=0.1U$.
We see that the effective and full model follow each other closely in Fig. \ref{fig:full}, and in Fig. \ref{fig:full zoomed} in a zoomed in cutout we see that the residual error decreases for increasing Fock-space cutoff of the dark mode. This truncation error is quantified in Fig. \ref{fig:Fock_convergence} where the full model is simulated with increasing cutoff compared to the effective model run at higher cutoff ($d_{eff}=12$, agreement to $10^{-10}$ with $d_{eff}=10$ and considered as converged).

\begin{figure}[!h]
    \centering
    \includegraphics[width=0.8\linewidth]{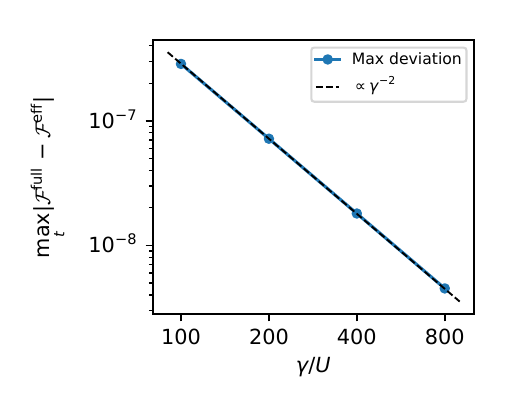}
    \caption{The maximum deviation over the period of one $XX$-flip between the full model and the Zeno-reduced model for a set of 4 values for $\gamma = 100U, 200U, 400U, 800U$ and cutoff $d_{dark}=6$ for the Fock space of the dark mode, with the effective model run at the same cutoff to identify Zeno-falloff. The dashed line shows a $\gamma^{-2}$ falloff. The fitted slope of the 4 values calculated falls as $\gamma^{-1.99}$.}
    \label{fig:func of gamma fidelity}
\end{figure}

To draw a conclusion on the validity of the Zeno-approximation we show how the mismatch between the effective and full model falls off as a function of $\gamma$ for a fixed dark mode cutoff $n_d$, which isolates the Zeno-error without including the Fock-truncation error of Fig. \ref{fig:Fock_convergence}. This is shown in Fig. \ref{fig:func of gamma fidelity}, and shows the falloff following the expected $\gamma^{-2}$ curve, which is a hallmark of the Zeno-approximation, and shows that at $\gamma =150U$, the smallest value used in the main body, the maximum mismatch between the full and effective model is (extrapolated) of order $\sim 10^{-7}$, confirming the validity of the Zeno-reduction used. This is also consistent with the estimate of Sec. \ref{deriving eff model}, where an estimate on the omitted terms of the Dyson-series was given. For $\gamma=200U$, the omitted terms contribute about $5.6\times10^{-4}$ \textit{relative} to the retained terms. The retained terms (the dissipators) at the current parameter choices contribute about $1.8\times10^{-4}$, so naively the expectation gives an absolute error of $\approx10^{-7}$ fron Zeno truncation. This is slightly larger than the observed value in Fig. \ref{fig:func of gamma fidelity}.  

\section{Derivation of Effective Lindbladian}
\label{Appendix Zeno}
The starting point is the expression for the time evolution operator (upon truncation)
\begin{align}
    \begin{split}
        U(\tau)_{\phi\phi} &= P_d + \tau P_d \mathcal{K}P_d\\
        &+ P_d\mathcal{K} \int_0^\tau \mathrm{d}\tau_1 \int_0^{\tau_1}\mathrm{d}\tau_2  e^{\mathcal{L}_d(\tau_1-\tau_2)}\mathcal{K} P_d.  \label{red prop appendix}
    \end{split}
\end{align}

The first two terms are fairly simple, so we restrict our attention to the third term and add the first two at the end. We work through the final term of Eq. \eqref{red prop appendix} from right to left, computing the product to the right of the exponent first.

\begin{equation}
    \begin{split}
    \mathcal{K}P_d \rho &= -\frac{i}{\gamma}\comm{\hat{H}_\phi}{\rhodp} -\frac{i}{\gamma}\frac{U}{N}\sum_{k\neq\phi}\sum_{\hat{T}\in\{\hat{A},\hat{B}\}}\bigg\{\hat{T}_k^\dagger\rhodp-\rhodp \hat{T}_k\bigg\} \\
    &- \frac{i}{\gamma}\frac{\lambda}{N}\sum_{k\neq\phi} \bigg\{ \hat{R}_k^\dagger\rhodp-\rhodp \hat{R}_k \bigg\}. \label{P_drho}
     \end{split}
\end{equation}
Where we have let $U(\tau)_{\phi\phi}$ act on $\rhodp$ for bookkeeping purposes. For brevity we use $\rho_{d\phi}$ as a shorthand notation for $\rho_{d}\otimes\rho_{\phi\phi}$.
For ease of notation we use the term $\hat{T}_k$ as a placeholder for $\hat{A}_k$ and $\hat{B}_k$, where these two are defined as
\begin{equation}
    \hat{A}_k = ((\alphadagop{\phi})^{2}-\zeta^{*2})\alphaop{k} \alphaop{2\phi-k}
\end{equation}
and likewise for $\hat{B}_k$ with $\alphaop{k}$ replaced by $\betaop{k}$, while we define $\hat{R}_k$ as
\begin{equation}
    \hat{R}_k = e^{-i(k-\phi)} (\betaop{k} \alphadagop{\phi} + \alphaop{k} \betadagop{\phi}),
\end{equation}
and $\hat{H}_\phi$ is the effective Hamiltonian from Eq. \eqref{effective hamiltonian}. We next calculate $\exp{(\mathcal{L}_d(\tau_1-\tau_2))}\hat{T}^\dagger_k\rhodp$ . From Zapletal \textit{et al.}~\cite[SM]{zapletal_stabilization_2022} the $\hat{A}_k$ term is calculated for a single chain, and we can generalize to include $\hat{B}_k$
\begin{equation}
    \mathcal{L}_d(\hat{A}_k^\dagger + \hat{B}_k^\dagger)\rhodp = -\frac{\gamma_k}{\gamma} (\hat{A}_k^\dagger + \hat{B}_k^\dagger)\rhodp,
\end{equation}
Correspondingly, we calculate $\exp{(\mathcal{L}_d(\tau_1-\tau_2))}\hat{R}^\dagger_k\rhodp$, where we include all steps for completeness. We first calculate
\begin{align}
\begin{split}
    &\mathcal{L}_d(\hat{R}_k^\dagger)\rhodp = \sum_{l\neq\phi}\frac{\gamma_l}{\gamma}\big(D[\alphaop{l}] + D[\betaop{l}]\big)\big(\hat{R}_k^\dagger\rhodp\big) \\
    &=\sum_{l\neq\phi}\frac{\gamma_l}{\gamma}\bigg(\alphaop{l}\big(\hat{R}_k^\dagger\rhodp\big)\alphadagop{l} -\frac{1}{2}\acomm{\alphadagop{l}\alphaop{l}}{\hat{R}_k^\dagger\rhodp} \\
    &+ \big[\alphaop{l}/\alphadagop{l}\leftrightarrow \betaop{l} / \betadagop{l}\big]\bigg).
\end{split}
\end{align}
We consider first only the terms containing $\alphaop{l}$ and $\alphadagop{l}$, the terms containing $\betaop{l}$ and $\betadagop{l}$ are treated in the same way. The first term in the second line is identically zero as $\rhodp\alphadagop{l} = 0$ for all $l\neq \phi$ from the condition set in Eq. \eqref{criteria operators}. This same condition holds for the second term in the anti-commutator, and we are left with 
\begin{align}
    \begin{split}
        &\sum_{l\neq\phi}\frac{\gamma_l}{\gamma}\bigg(\alphaop{l}\big(\hat{R}_k^\dagger\rhodp\big)\alphadagop{l} -\frac{1}{2}\acomm{\alphadagop{l}\alphaop{l}}{\hat{R}_k^\dagger\rhodp}\bigg)
        \\
        &=-\sum_{l\neq\phi}\frac{\gamma_l}{2\gamma}\alphadagop{l}\alphaop{l}\hat{R}_k^\dagger\rhodp \\
        &= -\sum_{l\neq\phi}\frac{\gamma_l}{2\gamma}e^{i(k-\phi)}\alphadagop{l}\alphaop{l}\big(\betadagop{k}\alphaop{\phi} + \alphadagop{k}\betaop{\phi}\big)\rhodp\\
        &= -\sum_{l\neq\phi}\frac{\gamma_l}{2\gamma}e^{i(k-\phi)}\alphadagop{l}\alphaop{l}\alphadagop{k}\betaop{\phi}\rhodp \\
        &=-\sum_{l\neq\phi}\frac{\gamma_l}{2\gamma}e^{i(k-\phi)}\alphadagop{l}\big(\delta_{lk}+\alphadagop{k}\alphaop{l}\big)\betaop{\phi}\rhodp \\
        &= -\frac{\gamma_k}{2\gamma}e^{i(k-\phi)}\alphadagop{k}\betaop{\phi}\rhodp.
    \end{split}
\end{align}
From the third to the fourth line, we commute $\alphaop{l}$ through the first term in the parenthesis and again use Eq. \eqref{criteria operators}, while from the fourth to the fifth we have a non-trivial commutator between $\alphaop{l}$ and $\alphadagop{k}$. We perform this to get $\alphaop{l}$ to the left hand side of $\rhodp$ to once again use Eq. \eqref{criteria operators} and resolve the sum. Doing the same calculation for the omitted $D[\betaop{l}]$terms yields
\begin{equation} 
    (\dots) = -\frac{\gamma_k}{2\gamma}e^{i(k-\phi)}\betadagop{k}\alphaop{\phi}\rhodp.
\end{equation}
Combining these two back together we get
\begin{equation}
    \mathcal{L}_d(\hat{R}_k^\dagger)\rhodp = -\frac{1}{2}\frac{\gamma_k}{\gamma} (\hat{R}_k^\dagger)\rhodp.
\end{equation}
In effect, this means we may replace all instances of $\mathcal{L}_d$ with its corresponding eigenvalue. This allows us to simply solve out the integrals.

\begin{align}
\begin{split}
    &\int_0^\tau \int_0^{\tau_1}d\tau_1d\tau_2 e^{\mathcal{L}_d(\tau_1-\tau_2)}\mathcal{K}P_d\rho = -\frac{i}{\gamma}\frac{\tau^2}{2}\comm{\hat{H}_\phi}{\rhodp}\\
    &-i\sum_{k\neq\phi}\sum_{\hat{T}\in\{\hat{A},\hat{B}\}}C_k\bigg\{\hat{T}^\dagger_k\rhodp-\rhodp \hat{T}_k\bigg\} \nonumber
    \\
    &- i\sum_{k\neq\phi} D_k \bigg\{\hat{R}_k^\dagger\rhodp-\rhodp \hat{R}_k\bigg\} \label{solve integrals}
    \end{split}
\end{align}
with 
 \begin{equation}
     C_k = \frac{U}{N}\bigg\{\frac{\tau}{\gamma_k} + \frac{\gamma}{\gamma_k^2}\bigg(e^{-\frac{\gamma_k}{\gamma}\tau}-1\bigg)\bigg\} \label{C_k}
 \end{equation}
and $D_k$ is $C_k$ with $U\to \lambda$ and $\gamma_k\to \frac{1}{2}\gamma_k$. Acting on this  from the left with $P_d \mathcal{K}$

\begin{align}
    \begin{split}
      &P_d\mathcal{K} \int_0^\tau \int_0^{\tau_1}d\tau_1d\tau_2 e^{\mathcal{L}_d(\tau_1-\tau_2)}\mathcal{K}P_d\rhodp \\ &= (\frac{-i}{\gamma})^2 \frac{\tau^2}{2}P_d\comm{\hat{H}}{\comm{\hat{H}_\phi}{\rho_{d\phi}}} \\
   &-\frac{1}{\gamma}\sum_{k\neq \phi}\sum_{\hat{T}\in\{\hat{A},\hat{B}\}}C_k\{P_d\comm{\hat{H}}{\hat{T}^\dagger_k\rhodp} +h.c.\} \\
   &-\frac{1}{\gamma}\sum_{k\neq\phi}D_k \{P_d\comm{\hat{H}}{\hat{R}^\dagger_k\rhodp} + h.c.\}. \label{Final step}
    \end{split}  
\end{align}
We start by considering the term in the  third line in Eq. \eqref{Final step}, namely the projection of the commutator between the Hamiltonian and the $\hat{T}_k^\dagger$ terms and their hermitian conjugates. Using linearity of the commutator and the results from~\cite{zapletal_stabilization_2022} we are left only with the contribution arising from the independent terms
\begin{align} 
\begin{split}
    &-\frac{2U}{N\gamma}\sum_{k\neq\phi}C_k\sum_{\hat{Z}_\phi\in\{\hat{A}_\phi,\hat{B}_\phi\}}\{\comm{\hat{Z}_\phi^\dagger}{\hat{Z}_\phi \rho_{\phi\phi}} + h.c.\}  \\
    &=\tau\frac{\Gamma}{\gamma}\sum_{\hat{Z}_\phi\in\{\hat{A}_\phi,\hat{B}_\phi\}}D[\hat{Z}_\phi] \rho_{\phi\phi}, \label{A12}
    \end{split}
\end{align}
For this term, the coupling Hamiltonian does not contribute when projected onto the two-mode subspace as seen by the identical result in the uncoupled case by Zapletal \textit{et al.}~\cite{zapletal_stabilization_2022}. The constants have been grouped together as
\begin{equation}
    \Gamma = 4\frac{U^2}{N^2}\sum_{k\neq \phi}\frac{1}{\gamma_k},
\end{equation}
where we have expanded the exponential in Eq. \eqref{C_k} to first order in $1/\gamma_k$.
For $\hat{Z}_\phi \in \{\hat{A}_\phi, \hat{B}_\phi\}$ we have
\begin{equation}
    \hat{Z}_\phi = \hat{z}_\phi^2-\zeta^2, \hat{z}_\phi\in\{\alphaop{\phi},\betaop{\phi}\}. \label{Z_phi twomode}
\end{equation}
 For the term in the last line of Eq. \eqref{Final step}, we get a non-zero contribution from the coupling term which results in an induced one-photon loss in both the effective modes in this model.

\begin{align}
\begin{split}
 &-\frac{1}{\gamma}\sum_{k\neq\phi}D_k \{P_d\comm{\hat{H}}{\hat{R}^\dagger_k\rhodp} + h.c.\} \\
 &= -\frac{\lambda}{N\gamma}\sum_{k\neq\phi} D_k \sum_{\hat{z}_\phi\in\{\alphaop{\phi},\betaop{\phi}\}}\{\comm{\hat{z_\phi}^\dagger}{\hat{z_\phi\rho_{\phi\phi}}} +h.c.\} \\
 &=\tau\frac{\chi}{\gamma}\sum_{\hat{z}_\phi\in\{\alphaop{\phi},\betaop{\phi}\}}D[\hat{z}_\phi]\rho_{\phi\phi}, \label{A15}
 \end{split}
\end{align}
where 
\begin{equation}
    \chi= 4\frac{\lambda^2}{N^2} \sum_{k\neq\phi} \frac{1}{\gamma_k},
\end{equation}
and we have, again, expanded the exponential for $D_k$. We note that the single photon decays at half the rate of the pair-decay, but in the effective rate this is exactly canceled by only having a single reabsorption channel, leaving the effective rates of two-photon loss and one-photon identical up to the difference in coupling strength; $\Gamma/\chi=U^2/\lambda^2$.  For the first term on the RHS of Eq. \eqref{Final step} we can use $\Tr_d\bigg[\comm{\hat{H}}{\rho_d\otimes \circ}\bigg] = \comm{\hat{H}_\phi}{\circ}$~\cite[SM]{zapletal_stabilization_2022} to resolve it. We return to the first two terms of Eq. \eqref{red prop appendix}, which are respectively $P_d$ and $(\tau/\gamma)\comm{\hat{H}_\phi}{\circ}$. While we calculated the last term of Eq. \eqref{red prop appendix} in the previous derivations, we kept $\rhodp$ as a bookkeeping device, and we now add it to the full Eq. \eqref{red prop appendix}, which becomes

\begin{align}
    \begin{split}
        U(\tau)_{\phi\phi}\rhodp &= P_d\rhodp - i\frac{\tau}{\gamma}\comm{\hat{H}_\phi}{\rhodp} \\
        &-\frac{\tau^2}{2\gamma^2}\comm{\hat{H}_\phi}{\comm{\hat{H}_\phi}{\rhodp}}+D[\circ]\rhodp.
    \end{split}
\end{align}
The final term is shorthand for the contribution from both the dissipative terms from Eqs. \eqref{A12} and \eqref{A15}, which both only act on the $\phi$ subspace. Tracing out the fixed vacuum modes with $\Tr_d$ leaves us with
\begin{align}
    \begin{split}
        &\Tr_d\big[U(\tau)_{\phi\phi}\rhodp \big] = U(\tau)_{\phi\phi}\rho_{\phi\phi} = \rho_{\phi\phi}(\tau) \\
        &=\rho_{\phi\phi} - i\frac{\tau}{\gamma}\comm{\hat{H}_\phi}{\rho_{\phi\phi}} -\frac{\tau^2}{2\gamma^2}\comm{\hat{H}_\phi}{\comm{\hat{H}_\phi}{\rho_{\phi\phi}}} \\
        &+D[\circ]\rho_{\phi\phi}. \label{a19}
    \end{split}
\end{align}
We now use the formal definition of the derivative $\partial\rho /\partial \tau = \lim_{\tau\to 0} \big(\rho(\tau) -\rho(0)\big)/\tau$. Inserting $\rho_{\phi\phi}(\tau)$ from Eq. \eqref{a19} into the formal definition, we see that the first term cancels and the second order term tends to zero. What we are left with is
\begin{equation}
    \frac{\partial\rho_{\phi\phi}}{\partial\tau} = -\frac{i}{\gamma}\comm{\hat{H}_\phi}{\rho_{\phi\phi}} + D[\circ]\rho_{\phi\phi}
\end{equation}
Switching variables back from $\tau$ to $t = \tau/\gamma$  cancels the $\gamma$ (recall there is an explicit $1/\gamma$ in both the dissipative terms as well),  and adding in the explicit form of the dissipative terms we can in the end rewrite our results in the form of a Lindblad master equation for the two $\phi$-modes in the Zeno limit as, done for the decoupled case by Zapletal \textit{et al.}\cite{zapletal_stabilization_2022}.
\begin{align}
\begin{split}
    \frac{\partial \rho_{\phi}(t)}{\partial t}  &= -i\comm{\hat{H}_\phi}{\rho_{\phi\phi}(t)} 
    +\Gamma\sum_{\hat{z}_\phi\in\{\alphaop{\phi},\betaop{\phi}\}} D[\hat{Z}_\phi]\rho_{\phi\phi}(t) \\
    &+ \chi \sum_{\hat{z}_\phi\in\{\alphaop{\phi},\betaop{\phi}\}} D[\hat{z}_\phi]\rho_{\phi\phi}(t),
    \end{split}
\end{align}
where $\hat{H}_\phi$ is the effective Hamiltonian for two modes defined in Eq. \eqref{effective hamiltonian}.
\section{The Projection onto $XX$}
\label{Projection XX}
Consider the coupling Hamiltonian
\begin{equation}
    \hat{H}_{BS}=\frac{\lambda}{N}(\alphadagop{\phi}\betaop{\phi}+h.c.)
\end{equation}
that we wish to project onto the logical manifold spanned by $\ket{0/1}_L = \ket{C_\pm}$. We define the projector as $P_L = P_A \otimes P_B$ with $P_A = \ket{C_+}\bra{C_+} + \ket{C_-}\bra{C_-}$, and likewise for $P_B$. We may factorize the projection
\begin{equation}
    P_L\hat{H}_{BS}P_L = \frac{\lambda}{N}\big[(P_A\alphadagop{\phi}P_A)(P_B\betaop{\phi}P_B)    + h.c. \big].
\end{equation}
We use the general properties of cat states $\alphaop{\phi}\ket{C_+} = \zeta r\ket{C_-}$ and $\alphaop{\phi}\ket{C_-} = \zeta r^{-1}\ket{C_+}$ to solve out this projection. $\zeta$ is defined in terms of the current parameters after Eq. \eqref{steady state form} and $r$ is the ratio of the cat normalization factors from Eq. \eqref{r def}. We calculate the projection of the raising and lowering operators as 
\begin{align}
\begin{split}
    P_A\alphaop{\phi}P_A &= \zeta(r\ket{C_-}\bra{C_+} + r^{-1}\ket{C_+}\bra{C_-})\\
    &= \zeta(uX - ivY)
    \end{split}\\
    \begin{split}
    P_A\alphadagop{\phi}P_A &= \zeta^*(r\ket{C_+}\bra{C_-} + r^{-1}\ket{C_-}\bra{C_+})\\
    &= \zeta^*(uX + ivY)
    \end{split}
\end{align}
where 
\begin{align}
    u &= \frac{r + r^{-1}}{2} \\
    v &= \frac{r-r^{-1}}{2}
\end{align}
and 
\begin{align}
    \ket{C_+}\bra{C_-} &= \frac{1}{2}(X+iY) \\
    \ket{C_-}\bra{C_+} &= \frac{1}{2}(X-iY).
\end{align}
For the full coupling Hamiltonian this gives
\begin{equation}
    P_L\hat{H}_{BS}P_L = \frac{2\lambda \abs{\zeta}^2}{N} (u^2X_AX_B + v^2 Y_AY_B). \label{projection omega}
\end{equation}
In the ideal limit for large average excitation number (\textit{i.e.} $r\to1$, $u\to 1$ and $v\to 0$) we have $\frac{\Omega}{2}X_AX_B$ with 
\begin{equation}
    \Omega = \frac{4\lambda\abs{\zeta}^2}{N} = \frac{4\lambda G}{U}
\end{equation}
and $T_{XX} = \pi/\Omega = \frac{\pi U}{4\lambda G}$ which is the term used in the main body of text in Eq. \eqref{time one flip}. Since the states have finite size, we instead retain all terms. We may factorize the evolution
\begin{equation}
    U = e^{-i\theta_X X_AX_B}e^{-i\theta_Y Y_AY_B}
\end{equation}
where the rotation angles are
\begin{align}
    \theta_X &= \frac{2\lambda\abs{\zeta}^2}{N} u^2 T_{XX} = \frac{\pi}{2}u^2 \\ 
    \theta_Y &= \frac{2\lambda\abs{\zeta}^2}{N} v^2 T_{XX} = \frac{\pi}{2}v^2.
\end{align}
The fidelity $\mathcal{F}(t=T_{XX}) = \abs{\bra{C_-C_+}U\ket{C_+C_-}}^2$ may be calculated analytically by noting
\begin{align}
\begin{split}
     \bra{C_-C_+}U\ket{C_+C_-} &= -i(\cos{\theta_X}\sin{\theta_Y}+   \sin{\theta_X}\cos{\theta_Y}) \\
     &= -i \sin{(\theta_X+\theta_Y)}
     \end{split}
\end{align}
so that
\begin{equation}
    \mathcal{F}(t=T_{XX}) = \sin^2{\big[\frac{\pi}{2}(u^2+v^2)\big]}. \label{fidelity almost final}
\end{equation}
Again, in the ideal limit of infinite average photon number, we regain unity fidelity by setting $u=1$ and $v=0$. We consider
\begin{equation}
    u^2+v^2 = \frac{r^2+r^{-2}}{2} = \frac{1+e^{-4\abs{\zeta}^2}}{1-e^{-4\abs{\zeta}^2}} 
\end{equation}
Expanding as a MacLaurin series in terms of $x = e^{-4\abs{\zeta}^2}$
\begin{equation}
    \frac{1+x}{1-x} = 1 +2x + O(x^2),
\end{equation}
we reinsert into Eq. \eqref{fidelity almost final} at lowest order to get
\begin{equation}
    \mathcal{F}(t=T_{XX}) \approx \sin^2{(\frac{\pi}{2} +\pi x)} = \cos^2{\pi x}.
\end{equation}
As we chose to focus on infidelity, we obtain 
\begin{equation}
    1-\mathcal{F}(t=T_{XX}) = 1-\cos^2{\pi x} = \sin^2{\pi x} \approx \pi^2x^2.
\end{equation}
Inserting back for $x$, this equation gives a floor on the attainable infidelity which is independent of $\lambda$ and fully determined by the average excitation $\expval{\hat{n}}$, given by 
\begin{equation}
    (1-\mathcal{F}(t=T_{XX}))_{\expval{\hat{n}}} \approx \pi^2 e^{-8\expval{\hat{n}}}.
\end{equation}
In this final step we reinserted $\abs{\zeta}^2 = \expval{\hat{n}} = NG/U$.

\bibliography{references,Manual_bib}

\end{document}